\documentclass[twocolumn]{aa}

\def\kms{$\rm km\;s^{-1}$}
\def\hour{$^{h}$}

\def\ha{H$\alpha$}
\def\h2{H$_{2}$}

\def\nii{[N~{\scriptsize II}]}

\def\ng{[N~{\scriptsize II}]$\,\lambda6583.41$}
\def\sii{[S~{\scriptsize II}]}

\def\oi{[O~{\scriptsize I}]}
\def\nai{Na~{\scriptsize I}}
\def\msun{M$_{\odot}$}

\def\maq{$\rm mag\;arcsec^{-2}$}
\def\mlr{(M/L$_{r})_{\odot}$}
\def\mlR{(M/L$_{R})_{\odot}$}
\def\msunpc{$\rm M_{\odot}\;pc^{-3}$}

\begin{document}

\thesaurus{ 03( 11.09.1: NGC~2179, NGC~2775; 
		11.11.1;       		
	    	11.19.2;          	
	    	11.19.6)}         	

\title{Dark matter in early-type spiral galaxies: the case of NGC~2179 and
of NGC~2775\thanks{Based on observations carried out at ESO, La Silla
(Chile) (ESO N. 52, 1-020) and on observations obtained with the VATT:
the Alice P. Lennon Telescope and the Thomas J. Bannan Astrophysics
Facility.}$^{\bf ,}$\thanks{Tables 4 to 42 are only available in
electronic form at the CDS via anonymous ftp to cdsarc.u-strasbg.fr
(130.79.128.5) or via http://cdsweb.u-strasbg.fr/Abstract.html.}}

\author{
	E.M.~Corsini,       	\inst{1},
	A.~Pizzella         	\inst{2}, 
	M.~Sarzi            	\inst{3},
	P.~Cinzano          	\inst{3}, 
	J.C.~Vega~Beltr\'an	\inst{4},
	J.G.~Funes, S.J.	\inst{3},
	F.~Bertola          	\inst{3},
	M.~Persic	    	\inst{5},
	and P.~Salucci		\inst{6}}

\offprints{E.M. Corsini}
\mail{{\tt corsini@pd.astro.it}}

\institute{
Osservatorio Astrofisico di Asiago, Dipartimento di Astronomia, 
Universit{\`a} di Padova, via dell'Osservatorio~8, I-36012 Asiago, Italy \and
European Southern Observatory, Alonso de Cordova 3107,   
Casilla 19001, Santiago 19, Chile \and
Dipartimento di Astronomia, Universit{\`a} di Padova, 
vicolo dell'Osservatorio~5, I-35122 Padova, Italy \and
Osservatorio Astronomico di Padova, Telescopio Nazionale Galileo, 
vicolo dell'Osservatorio~5, I-35122 Padova, Italy \and
Osservatorio Astronomico di Trieste, via G.B. Tiepolo 11, I-34131 Trieste, 
Italy \and
SISSA, via Beirut 2-4, I-34013 Trieste, Italy}

\date{Received..................; accepted...................}

\authorrunning{Corsini et al.} 
\titlerunning{Dark matter in Sa galaxies: NGC~2179 and NGC~2775}

\maketitle

\begin{abstract}

We present the stellar and ionized-gas velocity cur\-ves and
velocity-dispersion profiles along the major axis for six early-type
spiral galaxies.

Two of these galaxies, namely NGC~2179 and NGC~2775, are particularly
suited for the study of dark matter halos. Using their luminosity
profiles and modeling their stellar and gaseous kinematics, we derive
the mass contributions of the luminous and the dark matter to the
total potential. In NGC~2179 we find that the data (measured out to
about the optical radius $R_{\rm opt}$) unambiguously require the
presence of a massive dark halo.  For the brighter and bigger object
NGC~2775, we can rule out a significant halo contribution at radii $R
\la 0.6 \,R_{\rm opt}$. Although preliminary, these results agree with 
the familiar mass distribution trend known for late-type spirals of
comparable mass.


\begin{table*}[ht] 
\caption[]{Parameters of the sample galaxies.}
\begin{flushleft}
\begin{footnotesize}
\begin{tabular}{lllllllllllllc}
\hline
\noalign{\smallskip}
\multicolumn{1}{c}{object} &
\multicolumn{2}{c}{type} &
\multicolumn{1}{c}{$B_T$} &
\multicolumn{1}{c}{P.A.} &
\multicolumn{1}{c}{$i$} &
\multicolumn{1}{c}{$V_{\odot}$} &
\multicolumn{1}{c}{$V_0$} &
\multicolumn{1}{c}{$D$} &
\multicolumn{1}{c}{scale} &
\multicolumn{1}{c}{$R_{25}$} &
\multicolumn{1}{c}{$R_{\rm star}^{\rm far}$} &
\multicolumn{1}{c}{$R_{\rm gas}^{\rm far}$} &
\multicolumn{1}{c}{time} \\
\multicolumn{1}{c}{[name]} &
\multicolumn{1}{c}{[RSA]} &
\multicolumn{1}{c}{[RC3]} &
\multicolumn{1}{c}{[mag]} &
\multicolumn{1}{c}{[\degr]} &
\multicolumn{1}{c}{[\degr]} &
\multicolumn{1}{c}{[\kms]} &
\multicolumn{1}{c}{[\kms]} &
\multicolumn{1}{c}{[Mpc]} &
\multicolumn{1}{c}{[pc$''^{-1}$]} &
\multicolumn{1}{c}{[$''$]} &
\multicolumn{1}{c}{[$''$]} &
\multicolumn{1}{c}{[$''$]} &
\multicolumn{1}{c}{[\hour]} \\
\multicolumn{1}{c}{(1)} &
\multicolumn{1}{c}{(2)} &
\multicolumn{1}{c}{(3)} &
\multicolumn{1}{c}{(4)} &
\multicolumn{1}{c}{(5)} &
\multicolumn{1}{c}{(6)} &
\multicolumn{1}{c}{(7)} &
\multicolumn{1}{c}{(8)} &
\multicolumn{1}{c}{(9)} &
\multicolumn{1}{c}{(10)} &
\multicolumn{1}{c}{(11)} &
\multicolumn{1}{c}{(12)} &
\multicolumn{1}{c}{(13)} &
\multicolumn{1}{c}{(14)} \\
\noalign{\smallskip}
\hline
\noalign{\smallskip}
NGC~2179&Sa   &.SAS0..&13.22&170&51&2885$\pm$10&2673&35.6&172.6& 51& 41& 54&
8.5\\
NGC~2775&Sa(r)&.SAR2..&11.03&163&44&1350$\pm$10&1180&15.7& 76.1&128&100& 77&
6.5\\
NGC~3281&Sa   &.SAS2P*&12.70&138&69&3380$\pm$10&3098&41.3&200.2& 99& 86& 50&
6.5\\
IC~724&Sa   &.S..1..&13.4 & 60&55&5974$\pm$10&5853&78.0&378.2& 70& 62& 56&
2.0\\
NGC~4698&Sa   &.SAS2..&11.46&170&70& 992$\pm$10& 909&12.1& 58.7&119& 76&113&
4.7\\
NGC~4845&Sa   &.SAS2./&12.10& 75&72&1084$\pm$10& 980&13.1& 63.5&150& 87&100&
4.0\\
\noalign{\smallskip}
\hline
\noalign{\smallskip}
\noalign{\smallskip}
\noalign{\smallskip}
\end{tabular}
\begin{minipage}{18cm}
NOTES -- Col.(2): classification from RSA (Sandage \& Tamman 1981).
Col.(3): classification from RC3 (de Vaucouleurs et al. 1991).
Col.(4): total observed blue magnitude from RC3 except for IC~724 (RSA).
Col.(5): observed position angle. 
Col.(6): inclination from Rubin et al. (1985) except for NGC~2179 
(Tully 1988).
Col.(7): heliocentric velocity of the galaxy derived as center of symmetry
of the gas RC.
Col.(8): systemic velocity derived from $V_\odot$ corrected for the motion 
of the Sun with respect of the Local Group by $\Delta V=300\cos{b}\sin{l}$.
Col.(9): distance obtained as $V_0/H_0$ with $H_0=75$ \kms\ Mpc$^{-1}$.
Col.(11): radius of the 25 $B-$mag arcsec$^{-2}$ isophote from RC3.
Col.(12): radius of the farthest measured stellar velocity.
Col.(13): radius of the farthest measured gas velocity.
Col.(14): total integration time of the spectroscopic observation.
\end{minipage}
\end{footnotesize}
\end{flushleft}
\end{table*}

\end{abstract}


\section{Introduction}

Recent analyses of extended rotation curves (RCs) of late-type spiral
galaxies (Persic, Salucci \& Stel 1996) have confirmed that in spirals
of all luminosities a substantial dark matter (DM) component is
detectable already in the optical region. The effect is stronger at
lower luminosities: the dark-to-visible mass ratio at the optical
radius $R_{\rm opt}$\footnote{ $R_{\rm opt} = 3.2\, R_D$ is the
radius encompassing the $83\%$ of the total integrated light.  $R_D$
is the scale-length of the exponential surface brightness distribution
$I(r)=I(0)\, e^{-r/R_D}$.  For a Freeman (1970) disk, $R_{\rm opt}$
corresponds to the de Vaucouleurs 25 $B$-\maq\ photometric radius.}
scales with luminosity $\propto L^{-0.7}$.

For early-type spirals the status of our knowledge is different. The
RCs presently available for these objects are fragmentary (in
particular in the nuclear regions), and only extend to $\la 2\,R_D$
(see Rubin et al. 1985). Consequently, detailed mass decomposition
have so far not been possible for these systems.  In particular, it is
not known whether dark halos are unambiguously present also in
early-type spirals.

It may be conjectured that for a given $V(R_{\rm opt})$ the DM
fraction within the optical size is smaller in early than in late-type
spirals: this, because in an early spiral the conspicuous stellar
bulge, with $(M/L)_{\rm bulge} \ga 3 (M/L)_{\rm disk}$, can supply a
mass compact enough to make the rotation velocity higher than (see
Rubin et al. 1985), and the velocity profile different from, that of a
late spiral of similar luminosity.  In this case, the derivation of
the halo parameters would be more uncertain for early than for late
types: in fact, at small radii not two mass components (disk + halo,
like in Sc-Sd galaxies), but three mass components (bulge + disk +
halo) will have locally similar (solid-body like) behaviors. So for
non-extended RC data the mass solution of an Sa galaxy would be
degenerate even within the maximum-disk solution.

In this paper we present the velocity and velocity-dispers\-ion profiles
of the stars and the ionized gas, measured along the major axis, for
six early-type spirals. The six selected galaxies
(Table~1) were already known to show emission lines and
their photometric properties were known. Of these, 5 had already been
observed spectroscopically by Rubin et al. (1985), who obtained the
RCs of the ionized gas, and photometrically by Kent (1988). 
NGC~2179 was the only galaxy in our sample still lacking
spectroscopical and photometrical observations. To the originally
observed sample belonged the early-type spiral \object{NGC~3593}
too. Its stellar and the gaseous kinematics, fo\-und to exhibit a star
vs. star counterrotation, is presented and discussed by Bertola et
al. (1996).  Three-component models (bul\-ge + disk + halo) based on
observed photometry and kinematics are obtained for two galaxies of
the sample: NGC~2179 and NGC~2775.

\section{Observations and data reduction}

\subsection{The spectroscopic observations}

The spectroscopic observations of our sample galaxies were carried out
at the ESO 1.52-m Spectro\-sco\-pic Te\-le\-sco\-pe at La Silla on
February 15-19, 1994.

The telescope was equipped with the Boller \& Chivens Spe\-ctro\-graph.
The No.~26 grating with 1200 $\rm grooves~mm^{-1}$ was used in the
first order in combination with a $2\farcs5 \times 4\farcm2$ slit.  It
yielded a wavelength coverage of 1990~\AA\ between about 5200~\AA\ and
about 7190~\AA\ with a reciprocal dispersion of 64.80 
\AA~mm$^{-1}$. The instrumental resolution was derived measuring after
calibration the FWHM of 22 individual emission lines distributed all
over the spectral range in the 10 central rows of a comparison
spectrum. We checked that the measured FWHM's did not depend on
wavelength, and we found a FWHM mean value of 2.34~\AA. This
corresponds to $\sigma = 0.99$~\AA\ (i.e., $51$~\kms\ at 5800~\AA\ and
$46$~\kms\ at 6400~\AA).  The adopted detector was the No.~24
2048$\times$2048 Ford CCD, which has a $15\times15~\mu$m$^{2}$ pixel
size.  After an on-chip binning of 3 pixels along the spatial
direction, each pixel of the frame corresponds to 0.97~\AA\
$\times\;2\farcs43$.

The long-slit spectra of all the galaxies were taken along their
optical major axes.  At the beginning of each exposure the galaxy was
centered on the slit using the guiding camera.  Repeated exposures
(typically of 3600 s each) did ensure several hours of effective
integration without storing up too many cosmic rays.  Some long-slit
spectra of 8 late-G or early-K giant stars, obtained with the same
instrumental setup, served as templates in measuring the stellar
kinematics.  Their spectral classes range from G8III to K4III
(Hoffleit \& Jaschek 1982).

The typical value of the seeing FWHM during the observing nights,
measured by the La Silla Differential Image Motion Monitor (DIMM), was
$1'' - 1\farcs5$.  Comparison helium-argon la\-mp exposures were taken
before and after every object exposure. The logs of the spectroscopic
observations of galaxies and template stars are reported in
Tables~2 and 3, respectively.

\begin{table}[tb] 
\caption{Log of spectroscopic observations (galaxies).}
\begin{flushleft}
\begin{tabular}{llcc}
\hline
\noalign{\smallskip} 
\multicolumn{1}{c}{object} & 
\multicolumn{1}{c}{date}&
\multicolumn{1}{c}{U.T.}& 
\multicolumn{1}{c}{time} \\
\multicolumn{1}{c}{[name]} & 
\multicolumn{1}{c}{[d m y]} &  
\multicolumn{1}{c}{[h m]} & 
\multicolumn{1}{c}{[s]} \\
\multicolumn{1}{c}{(1)} &
\multicolumn{1}{c}{(2)} &
\multicolumn{1}{c}{(3)} &
\multicolumn{1}{c}{(4)} \\    
\noalign{\smallskip} 
\hline
\noalign{\smallskip} 
\object{NGC~2179} & 15 Feb 1994 & 01 24 & 3600 \\
NGC~2179 & 15 Feb 1994 & 02 32 & 3600 \\
NGC~2179 & 16 Feb 1994 & 01 17 & 5400 \\
NGC~2179 & 17 Feb 1994 & 01 41 & 3600 \\
NGC~2179 & 18 Feb 1994 & 00 26 & 3600 \\
NGC~2179 & 18 Feb 1994 & 01 35 & 3600 \\
NGC~2179 & 19 Feb 1994 & 00 19 & 3600 \\
NGC~2179 & 19 Feb 1994 & 01 27 & 3600 \\
\noalign{\smallskip}
\object{NGC~2775} & 15 Feb 1994 & 04 10 & 3600 \\
NGC~2775 & 15 Feb 1994 & 05 17 & 3600 \\
NGC~2775 & 16 Feb 1994 & 03 56 & 5400 \\
NGC~2775 & 17 Feb 1994 & 02 47 & 3600 \\
NGC~2775 & 17 Feb 1994 & 03 53 & 3600 \\
NGC~2775 & 18 Feb 1994 & 02 41 & 3600 \\
\noalign{\smallskip}
\object{NGC~3281} & 16 Feb 1994 & 04 39 & 5400 \\
NGC~3281 & 16 Feb 1994 & 06 14 & 3600 \\
NGC~3281 & 17 Feb 1994 & 05 02 & 3600 \\
NGC~3281 & 18 Feb 1994 & 03 50 & 3600 \\
NGC~3281 & 19 Feb 1994 & 02 34 & 3600 \\
NGC~3281 & 19 Feb 1994 & 03 40 & 3600 \\
\noalign{\smallskip}
\object{IC~724}   & 17 Feb 1994 & 06 16 & 3600 \\
IC~724   & 17 Feb 1994 & 07 21 & 3600 \\
\noalign{\smallskip}
\object{NGC~4698} & 16 Feb 1994 & 07 22 & 3600 \\
NGC~4698 & 16 Feb 1994 & 08 28 & 3600 \\
NGC~4698 & 18 Feb 1994 & 07 17 & 3600 \\
NGC~4698 & 19 Feb 1994 & 07 42 & 6000 \\
\noalign{\smallskip}
\object{NGC~4845} & 15 Feb 1994 & 07 02 & 3600 \\
NGC~4845 & 15 Feb 1994 & 08 08 & 3600 \\
NGC~4845 & 17 Feb 1994 & 05 02 & 3600 \\
NGC~4845 & 18 Feb 1994 & 08 25 & 3600 \\
\noalign{\smallskip} 
\hline
\noalign{\smallskip}
\noalign{\smallskip}
\noalign{\smallskip}
\end{tabular}
\begin{minipage}{8cm}
NOTES -- Cols.(2-3): date and time of start of exposure.
Col.(4): exposure time.
\end{minipage}
\label{tab:log_g}
\end{flushleft}
\end{table}

\begin{table}[tb] 
\caption{Log of spectroscopic observations (template stars).}
\begin{flushleft}
\begin{tabular}{lllcc}
\hline
\noalign{\smallskip} 
\multicolumn{1}{c}{object} &
\multicolumn{1}{c}{type} & 
\multicolumn{1}{c}{date}&
\multicolumn{1}{c}{U.T.}& 
\multicolumn{1}{c}{time} \\
\multicolumn{1}{c}{[name]} &
\multicolumn{1}{c}{[BSC]} & 
\multicolumn{1}{c}{[d m y]} &  
\multicolumn{1}{c}{[h m]} & 
\multicolumn{1}{c}{[s]} \\    
\multicolumn{1}{c}{(1)} &
\multicolumn{1}{c}{(2)} &
\multicolumn{1}{c}{(3)} &
\multicolumn{1}{c}{(4)} &
\multicolumn{1}{c}{(5)} \\
\noalign{\smallskip} 
\hline
\noalign{\smallskip} 
\object{HR~2035}  & G8III   & 15 Feb 1994 & 00 53 &  2 \\
HR~2035  & G8III   & 17 Feb 1994 & 00 04 &  2 \\
HR~2035  & G8III   & 17 Feb 1994 & 00 07 &  3 \\
HR~2035  & G8III   & 17 Feb 1994 & 00 10 &  4 \\
HR~2035  & G8III   & 17 Feb 1994 & 00 16 &  3 \\
HR~2035  & G8III   & 19 Feb 1994 & 00 07 &  2 \\
HR~2035  & G8III   & 19 Feb 1994 & 00 09 &  3 \\
\noalign{\smallskip}
\object{HR~2429}  & K1III   & 18 Feb 1994 & 00 08 &  2 \\
HR~2429  & K1III   & 18 Feb 1994 & 00 11 &  2 \\
HR~2429  & K1III   & 18 Feb 1994 & 00 14 &  3 \\
\noalign{\smallskip}
\object{HR~2443}  & K1III   & 16 Feb 1994 & 00 25 &  3 \\
\noalign{\smallskip} 
\object{HR~2503}  & K4III   & 16 Feb 1994 & 00 35 &  4 \\
HR~2503  & K4III   & 16 Feb 1994 & 00 38 &  4 \\
HR~2503  & K4III   & 16 Feb 1994 & 00 41 &  4 \\
HR~2503  & K4III   & 16 Feb 1994 & 00 46 &  4 \\
HR~2503  & K4III   & 16 Feb 1994 & 00 49 &  8 \\
\noalign{\smallskip} 
\object{HR~3431}  & K4III   & 15 Feb 1994 & 03 52 &  4 \\
\noalign{\smallskip} 
\object{HR~5100}  & K0III   & 19 Feb 1994 & 09 29 & 10 \\
HR~5100  & K0III   & 19 Feb 1994 & 09 32 &  7 \\
\noalign{\smallskip} 
\object{HR~5196}  & K0.5III & 16 Feb 1994 & 09 33 &  4 \\
HR~5196  & K0.5III & 16 Feb 1994 & 09 38 & 15 \\
HR~5196  & K0.5III & 16 Feb 1994 & 09 43 & 10 \\
\noalign{\smallskip} 
\object{HR~5315}  & K3III   & 15 Feb 1994 & 09 14 &  4 \\
HR~5315  & K3III   & 15 Feb 1994 & 09 17 &  3 \\
HR~5315  & K3III   & 15 Feb 1994 & 09 21 &  3 \\
\noalign{\smallskip} 
\object{HR~5601}  & K0.5III & 15 Feb 1994 & 09 29 &  3 \\
HR~5601  & K0.5III & 15 Feb 1994 & 09 32 &  4 \\
HR~5601  & K0.5III & 15 Feb 1994 & 09 35 &  4 \\
HR~5601  & K0.5III & 15 Feb 1994 & 09 39 & 10 \\
\noalign{\smallskip} 
\hline
\noalign{\smallskip}
\noalign{\smallskip}
\noalign{\smallskip}
\end{tabular}
\begin{minipage}{8cm}
NOTES -- Col.(2): spectral class of the template star from The Bright Star 
Catalogue (Hoffleit \& Jaschek 1982). Cols.(3-4): date and time of start of
exposure. Col.(5): exposure time.
\end{minipage}
\end{flushleft}
\end{table}

\subsubsection{Data reduction}

Using standard MIDAS\footnote{MIDAS is developed and maintained by the
European Southern Observatory.} routines, all the spectra were bias
subtracted, flat-field corrected by quartz lamp exposures, and clea\-ned
from cosmic rays.  Cosmic rays were identified by comparing the counts
in each pixel with the local mean and standard deviation, and then
corrected by substituting a suitable value.

A small misalignment was present between the CCD and the slit.  We
measured a difference of $\Delta Y \simeq 1.4$ pixel between the
positions of the center of the stellar continuum near the blue and red
edge of the spectra.  When measuring the stellar kinematics, the tilt
had to be removed.  This was done by rotating the spectra by a
suitable angle ($\theta=0\fdg04$) before the wavelength
calibration. We noticed however that the sharp line profile of the
emission lines was spoiled by the rotating algorithm.  For this reason
no rotation was applied when measuring the ionized-gas kinematics.

The wavelength calibration was done using the MIDAS pa\-ckage XLONG. We
determined the velocity error possibly introduced by the calibration
measuring the `velocity curve' of a sample of 24 OH night-sky emission
lines distributed all over the spectral range.  The velocity did not
show any significant dependence on radius, indicating that the
wavelength rebinning had been done proper-Ely.  We found a mean
deviation from the predicted wavelengths (Osterbrock et al. 1996) of
$2$ \kms.

After calibration, the different spectra obtained for a given galaxy
were co-added using their stellar-continuum centers as reference.  For
each spectrum the center was assumed as the center of the Gaussian
fitting the mean radial profile of the stellar continuum.  The
contribution of the sky was determined from the edges of the resulting
galaxy frames and then subtracted.

\subsubsection{Measuring the gas kinematics}

The ionized-gas velocities ($v_g$) and velocity dispersions
($\sigma_g$) were measured by means of the MIDAS package ALICE. We
measured the \nii\ lines ($\lambda\lambda\,$6548.03, 6583.41 \AA), the
\ha\ line ($\lambda\,$6562.82 \AA), and the \sii\ lines
($\lambda\lambda\,$6716.47, 6730.85 \AA), where they were clearly
detected. The position, the FWHM, and the uncalibrated flux $F$ of
each emission line were determined by interactively fitting one
Gaussian to each line plus a polynomial to its local 
continuum. The center wavelength of the fitting Gaussian was
converted into velocity in the optical convention $v = cz$; then
the standard heliocentric correction was applied. The Gaussian FWHM
was corrected for the instrumental FWHM, and then converted into the
velocity dispersion $\sigma$. In the regions where the intensity of
the emission lines was low, we binned adjacent spectral rows in order
to improve the signal-to-noise ratio, $S/N$, of the lines.

\begin{figure} 
\vspace*{9cm}
\includegraphics{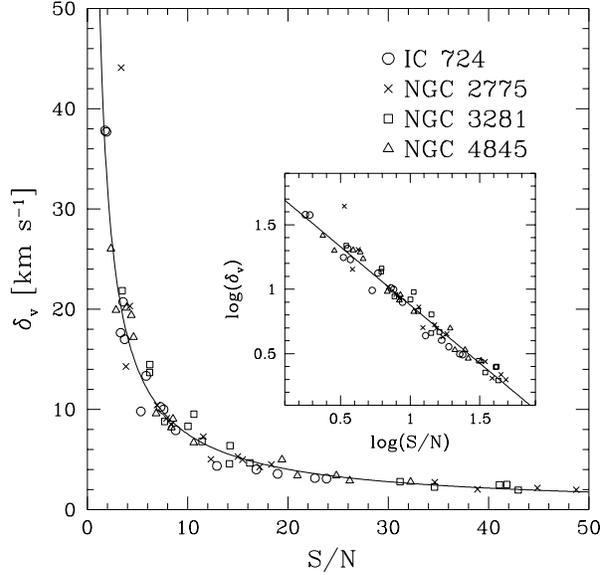}
\caption[]{The variation of the r.m.s. velocity error $\delta_v$ as a
  function of the signal-to-noise ratio $S/N$.  The symbols ({\it
  triangles, circles, squares\/} and {\it crosses\/}) represent the
  r.m.s. $\delta_v$ of the quadratic polynomials fitting the `velocity
  curves' of the same 16 night-sky emission lines selected in 4
  different spectra (NGC~2775, NGC~3281, IC~724 and NGC~4845 respectively) as
  a function of the relevant $S/N$ ratio of the lines.  The {\it full
  line\/} represents the power law fitting the data.  In the insert the
  data are plotted in logarithmic scale}
\label{fig:snratio}
\end{figure}

We expressed the variation of the r.m.s. velocity error $\delta_v$ as
a function of the relevant line $S/N$ ratio. In order to find the
expression for $\delta_v = \delta_v (S/N)$, we selected the same 16
night-sky emission lines in the spectra of NGC~2775, NGC~3281, IC~724 and
NGC~4845.  Such night-sky emissions were chosen to have different
intensities and different wavelengths between 6450~\AA\ and 6680~\AA\
(i.e. the wavelength range of the observed emission lines of the
ionized gas) in the four spectra.  We derived the sky spectra by
averaging several rows along the spatial direction in a galaxy-light
free region.  Using the above package, we interactively fitted one
Gaussian emission plus a polynomial continuum to each selected sky
line and its local continuum. We derived the flux $F$ and the FWHM of
the sample lines, taking the ratio $F/{\rm FWHM}$ as the signal
$S$. For each galaxy spectrum the noise $N$ was defined as the
r.m.s. of the counts measured in regions of the frame where the
contributions of both the galaxy and the sky lines were negligible.
The resulting $S/N$ range was large ($1 \leq S/N \leq 50$). For each
sample emission line we then measured, by means of an automatic
procedure, the night-sky `velocity curve' along the full slit
extension.  The wavelengths of the emissions were evaluated with
Gaussian fits and then converted to velocities. The radial profiles of
the sky-line velocities were then fitted by quadratic polynomials. We
assumed the r.m.s.  of the fit to each `velocity curve' to be the
$1\sigma$ velocity error. Fig.~\ref{fig:snratio} shows the good
agreement between the distributions of the $(S/N,\,\delta_v)$
measurements taken in the 4 different spectra.  In log-log scale, the
$(S/N,\delta_v)$ relation is well represented by a straight line, that
corresponds to:

\begin{equation} 
\delta_v ~=~ 60.4 \ \left( \frac{S}{N} \right)^{-0.90} {\rm km\ s}^{-1} 
\label{eq:sn} 
\end{equation} 

\noindent
(least-squares fit). This result agrees with Keel's (1996) relation
$\delta_v \propto (S/N)^{-1}$, based on numerical simulations.  Once
the relevant $S/N$ ratio of the emission had been derived, we obtained
$\delta_v$ for each velocity measurement of the ionized-gas component
by means of Eq.~\ref{eq:sn}.  The gas velocities derived independently
from different emission lines are in mutual agreement within their
errors $\delta_v$.

\begin{figure*}[ht] 
\vspace*{18cm}
\includegraphics{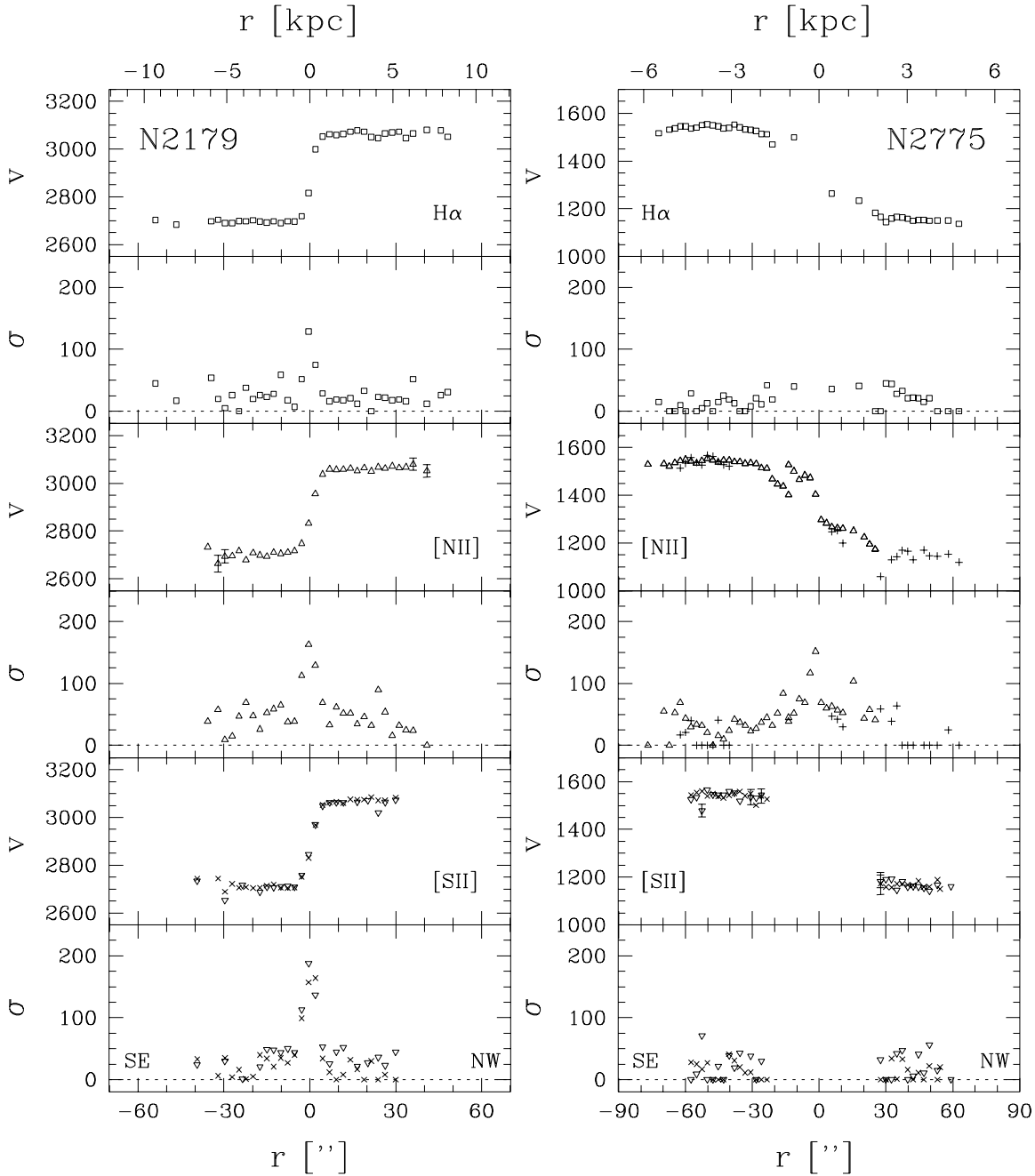}
\vspace*{.25cm}
\caption[]{NGC~2179 and NGC~2775 observed major-axis velocity curves
  and velocity dispersion profiles of ionized gas.  The kinematics
  measured from the \ha\ line is represented by {\it open squares\/} ($\sq$). 
  The {\it crosses\/} ($+$) and the {\it open triangles\/}
  ($\bigtriangleup$) refer to the data obtained from the \nii\
  ($\lambda\lambda\,$6548.03, 6583.41 \AA) The {\it times\/} ($\times$) and
  the {\it reversed triangles\/} ($\bigtriangledown$) represent kinematics
  desumed from the \sii\ ($\lambda\lambda\,$6716.47, 6730.85 \AA) lines.
  For the velocities only the errors greater than symbols are plotted}
\label{fig:gas_a}
\end{figure*}

\begin{figure*}[ht] 
\vspace*{18.5cm}
\includegraphics{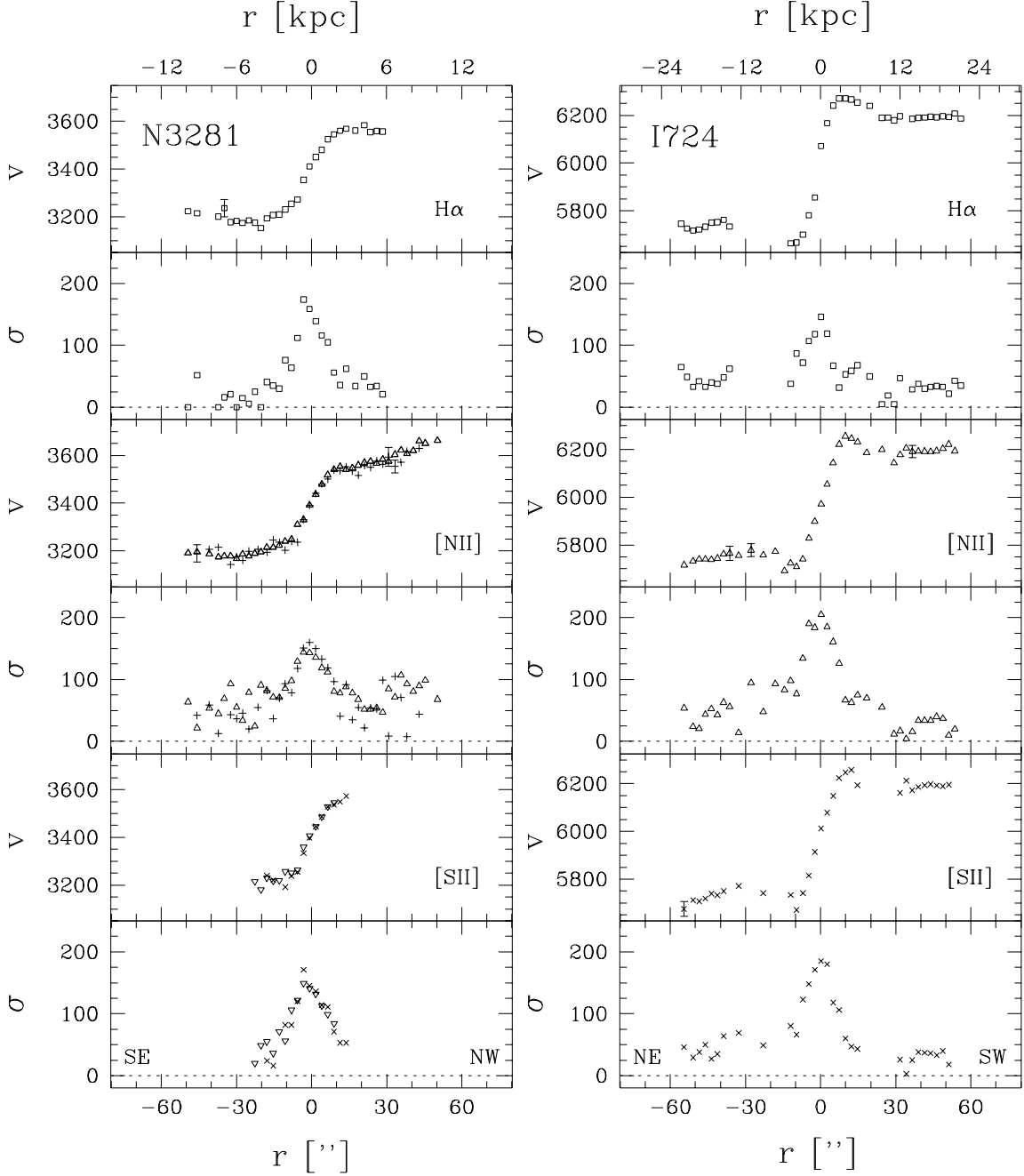}
\vspace*{.25cm}
\caption[]{Same as in Fig.~2, but for NGC~3281 and IC~724}
\label{fig:gas_b}
\end{figure*}

\begin{figure*}[ht] 
\vspace*{18.5cm}
\includegraphics{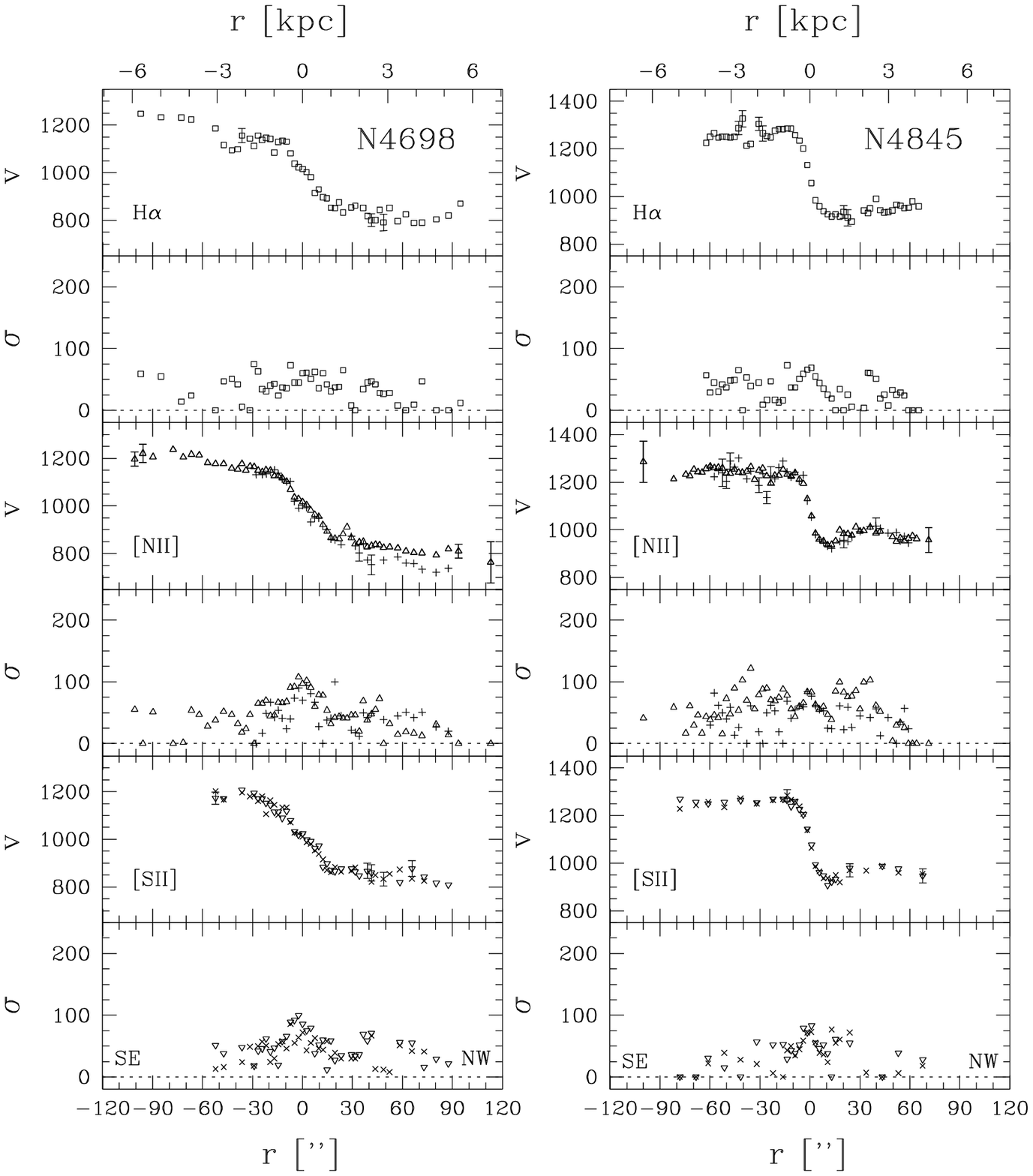}
\vspace*{.25cm}
\caption[]{Same as in Fig.~2, but for NGC~4698 and NGC~4845}
\label{fig:gas_c}
\end{figure*}

The ionized-gas velocities and velocity dispersions from \nii\
($\lambda\lambda\,$6548.03, 6583.41 \AA), \sii\
($\lambda\lambda\,$6716.47, 6730.85 \AA), and \ha\ are reported in:
Tables~4 -- 7 for NGC~2179; Tables~10 -- 14 for NGC~2775; Tables~17 --
21 for NGC~3281; Tables~24 -- 26 for IC~724; Tables~29 -- 33 for
NGC~4698; and Tables~36 -- 40 for NGC~4845.  Each table reports the
galactocentric distance $r$ in arcsec (Col.~1), the observed
heliocentric velocity $v$ and its error $\delta_v$ in \kms\ (Col.~2),
the velocity dispersion $\sigma$ in \kms\ (Col.~3), the number $n$ of
spectrum rows binned along the spatial direction (Col.~4), and the
signal-to-noise ratio $S/N$ of the emission line (Col.~5).  The \ha ,
\nii\ and \sii\ kinematics of the sample objects are plotted in
Figs.~\ref{fig:gas_a}~--~\ref{fig:gas_c}.

The final ionized-gas kinematics is obtained by averaging, at each
radius, the gas velocities and velocity dispersions derived
independently from the different emission lines. The gas velocity
($v_g$) and velocity error ($\delta v_g$) are respectively the
$1/\sigma_v^{2}$-weighted mean velocity and its uncertainty.  The gas
velocity dispersion ($\sigma_g$) and velocity-dispersion error
($\delta \sigma_g$) are the me\-an velocity dispersion and its
uncertainty. (No error is given when only one velocity dispersion
measurement is available.)

The kinematics of the ionized gas is reported in: Table~8 for
NGC~2179; Table~15 for NGC~2775; Table~22 for NGC~3281; Table~27 for
IC~724; Table~34 for NGC~4698; and Table~41 for NGC~4845.  Each table
reports the galactocentric distance $r$ in arcsec (Col.~1), the mean
heliocentric velocity $v_g$ and its error $\delta v_g$ in
\kms\ (Col.~2), the mean velocity dispersion $\sigma_g$ and its error 
$\delta \sigma_g$ in \kms\ (Col.~3). The ionized-gas velocity and
velocity-dispersion profiles are plotted in Fig.~\ref{fig:kin} 
for all our galaxies.

\subsubsection{Measuring the stellar kinematics}

The stellar velocities ($v_\star$) and velocity dispersions
($\sigma_\star$) of the sample galaxies were measured from the
absorption lines in the wavelength range between about 5200~\AA\ and
6200~\AA.  We used an interactive version of the Fourier Quotient
Method (Sargent et al.  1977) as applied by Bertola et al. (1984).
The K0III star HR~5100 was taken as template: it has a radial velocity
of $-0.9$ \kms\ (Wilson 1953) and a rotational velocity of $10$ \kms\
(Bernacca \& Perinotto 1970).

The spectra of the galaxies and the template star were rebinned to a
logarithmic wavelength scale, continuum subtract\-ed, and masked at
their edges by means of a cosine bell function of $20\%$ length.  At
each radius the galaxy spectrum was assumed to be the convolution of
the template spectrum with a Gaussian broadening function
characterized by the parameters $\gamma$, $v_\star$ and
$\sigma_\star$. They respectively represent the line strength of the
galaxy spectrum relative to the template's, and the line-of-sight
stellar velocity and velocity dispersion.  The parameters of the
broadening function, and consequently the stellar kinematics, were
obtained by a least-squares fitting in the Fourier space of the
broadened template spectrum to the galaxy spectrum in the wavenumber
range $\left[k_{min},k_{max}\right]=\left[5,440\right]$.  In this way
we rejected the low-frequency trends (corresponding to $k<5$) due to
the residuals of continuum subtraction and the high-frequency noise
(corresponding to $k>440$) due to the instrumental resolution. (The
wavenumber range is important in particular in the Fourier fitting of
lines with non-Gaussian profiles, see van der Marel \& Franx 1993 and
Cinzano \& van der Marel 1994).
In deriving the above kinematical properties, the regions $5569.0 <
\lambda < 5585.0$~\AA\ and $5884.0 < \lambda < 5900.0$~\AA\ were
masked because of contamination from bad subtraction of the night-sky
emission lines of \oi\ ($\lambda\,5577.34$~\AA) and \nai\
($\lambda\,5889.95$~\AA).

\begin{figure*}[ht] 
\vspace*{18.9cm}
\includegraphics{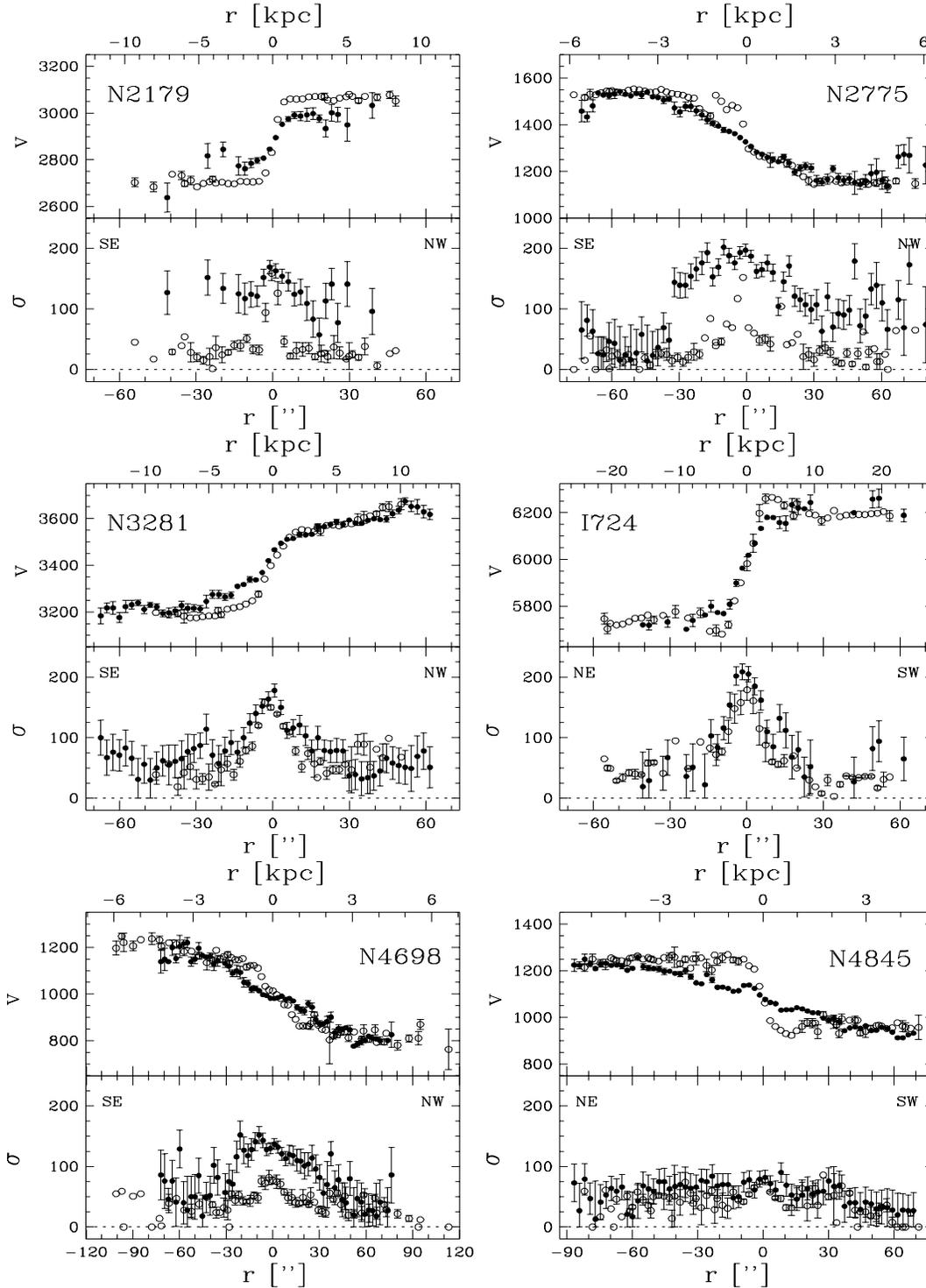}
\vspace*{.25cm}
\caption[]{Major-axis observed velocity curves and velocity dispersion
  profiles of the stellar ({\em filled circles\/}) and the gaseous 
  ({\it open circles\/}) components for the sample galaxies.
  For the velocities the errorbars smaller than symbols are not plotted}
\label{fig:kin}
\end{figure*}

The measured stellar kinematics is reported in: Table~9 for NGC~2179;
Table~16 for NGC~2775; Table~23 for NGC~3281; Table~28 for IC~724;
Table~35 for NGC~4698; and Table~42 for NGC~4845.  Each table reports
the galactocentric distance $r$ in arcsec (Col.~1), the heliocentric
velocity $v_\star$ and its error $\delta v_\star$ in \kms\ (Col.~2),
the velocity dispersion $\sigma_\star$ and its error $\delta
\sigma_\star$ in \kms\ (Col.~3).  The stellar velocity and
velocity-dispersion profiles are plotted in Fig.~\ref{fig:kin}. 

\subsection{The photometric observations}

The observation in the Cousin $R-$band of NGC~2179 was performed on
March 11, 1997 at the 1.83-m Vatican Advanced Technology Telescope
(VATT) at Mt. Graham International O\-bservatory. A back-illuminated
2048$\times$2048 Loral CCD with $15$$\times15~\mu$m$^{2}$ pixels was
used as detector at the aplanatic Gregorian focus, f/9. It yielded a
field of view of $6\farcm4\times6\farcm4$ with an image scale of
$0\farcs4$ pixel$^{-1}$ after a $2\times2$ pixel binning. The gain and
the readout noise were 1.4 e$^-$ ADU$^{-1}$ and 6.5 e$^-$
respectively.  We obtained 3 images of 120 s with the $R$ 3.48-inch
square filter.

The data reduction was carried out using standard IRAF\footnote{IRAF
is distributed by the National Optical Astronomy Observatories which
are operated by the Association of Universities for Research in
Astronomy (AURA) under cooperative agreement with the National Science
Foundation.} routines.  The images were bias subtracted and then
flat-field corrected.  They were shifted and aligned to an accuracy of
a few hundredths of a pixel using field stars as reference.  After
checking that the point spread functions (PSFs) in the images were
comparable, they were averaged to obtain a single $R$ image.  The
cosmic rays were identified and removed during the averaging
routine. A Gaussian fit to the intensity profile of field stars in the
resulting image allowed us to estimate a seeing PSF FWHM of
$1\farcs8$.

The sky subtraction and the elliptical fitting to the galaxy isophotes
were performed by means of the Astronomical Images Analysis Package
(AIAP: Fasano 1990).  The sky level was determined by a polynomial fit
to surface brightness in frame regions not contaminated by ga\-la\-xy
light; then it was subtracted out from the total signal. The isophote
fitting was performed masking the frame's bad columns and the bright
field stars.  We then obtained surface brightness, ellipticity,
major-axis position angle, and the $\cos 4\theta$ Fourier coefficient
of the isophote's deviations from elliptical as a function of radius
along the major axis.  No photometric standards were observed.  Thus
the absolute calibration was made using the photometric quantities
edited by Lauberts \& Valentijn (1989) in the same band.  We set the
surface brightness of an isophote with semi-major axis of
$a=20\farcs6$ to the value $\mu = 21.23$ $R$-\maq\ .

\section{The stellar and ionized gas kinematics}

The resulting kinematics of all our galaxies are shown in
Fig.~\ref{fig:kin}. The plotted velocities are as observed (no
inclination correction is applied).  In the following we briefly
discuss each individual object.  At each radius, $V_\star$ ($\equiv
|v_\star-V_\odot|$) and $V_g$ ($\equiv |v_g-V_\odot|$) are the
observed rotation velocities of the stars and the ionized gas,
respectively.\\

\medskip

\noindent
{\bf NGC~2179 --} The gas and stellar kinematics respectively extend to
$50''$ ($8.6$ kpc) and $40''$ ($6.9$ kpc) on either side of the
nucleus. Outwards of $40''$, the stellar and the gas radial velocities
are comparable.\\
{\it Stars.} In the inner $10''$ ($1.7$ kpc), $V_\star$ increases to
$\sim\,100$ \kms .  At the center $\sigma_\star\,\sim\,170$ \kms\ ;
away from the nucleus it remains high ($\sigma_\star \geq 100$ \kms),
and possibly even rises.\\
{\it Gas.} $V_g$ has a steeper gradient than $V_\star$, reaching a
value of $V_g=190$ \kms\ at $|r| \sim\,6''$ ($1.0$ kpc). $\sigma_g$ is
strongly peaked ($\sim\,150$ \kms ) at the center; at $|r| > 6''$ it
drops rapidly to $\sigma_g \simeq 30$ \kms .\\
A circumnuclear Keplerian disk of ionized gas has been recently
discovered in the center of NGC~2179 by Bertola et al. (1998a) by means
of optical ground-based observations. By modeling the motion of the
gaseous disk they inferred the presence of a central mass
concentration of $10^9$ \msun .\\

\noindent
{\bf NGC~2775 --} The gas and stellar kinematics is measured out to
$80''$ ($6.1$ kpc) from the center.\\
{\it Stars.} $V_\star$ increases almost linearly with radius, up to
about 130 \kms\ at $|r| \sim\,22''$ ($1.7$ kpc); for $22'' \leq |r| \leq
30''$ it remains approximately constant; further out it increases to
185 \kms\ and then flattens out. At $-30\arcsec \leq r \leq +20''$,
$\sigma_\star > 150$ \kms\ ; farther out it declines to $\sim\,40$
\kms\ and $\sim\,100$ \kms\ in the SE and NW side, respectively.\\
{\it Gas.} The gas behaves differently in the two regions
$|r|\leq20''$ ($1.5$ kpc) and $|r| > 20''$. The \ng\ line (the only
emission line detected in both regions, see Fig.~\ref{fig:gas_a})
shows the presence of two kinematically distinct gas components, named
component (i) and (ii).  Component (i) rotates with the same velocity
as the stars but with a lower velocity dispersion; component (ii)
rotates faster than the stars for $0 > r \ga -15''$ (1.1 kpc). Both
components show up simultaneously in the spectrum only at $|r| \simeq
14''$, where a double peak in the emission line is clearly
detected. $\sigma_g$ peaks (160 \kms) at the center and rapidly drops
to $\sim\,30$ \kms\ off center.\\

\noindent
{\bf NGC~3281 --} The stellar kinematics extends to $90''$ ($18$ kpc)
and $60''$ (12 kpc) in the SE and the NW side, respectively. The gas
kinematics can only be measured within $50''$ (10 kpc) on each side of
the nucleus.\\
{\it Stars.} The stars exhibit a rather shallow rotation gradient: at
$r = 10''$ ($2$ kpc), $V_\star \sim\,100$ \kms.  At the center
$\sigma_\star \sim\,180$ \kms\ ; off center it decreases to,
respectively, $\sim\,70$ \kms\ in the SE side and $\sim\,50$ \kms\ in
the NW side.\\
{\it Gas.} $V_g$ has a steep gradient, reaching 150 \kms\ at $10''$ 
and then 200 \kms\ at $20''$ (4 kpc). At the center $\sigma_g
\sim$160 \kms, while at $|r| > 20''$ it falls to $\sim$50 \kms;
$\sigma_g \sim \sigma_\star$ for $|r| < 10''$.\\

\noindent
{\bf IC~724 --} The stellar kinematics is observed out to $60''$
($22.7$ kpc) and $40''$ ($15.1$ kpc) in the SW and NE sides,
respectively. The ionized-gas kinematics extends to $60''$ on each
side of the nucleus. For $|r| > 20''$ ($7.6$ kpc), the gas and stellar
kinematics are similar. For $|r| < 20''$, $\sigma_g$ is centrally
peaked at $\sim\,180$ \kms , remaining lower than $\sigma_\star$.\\
{\it Stars.} $V_\star$ increases linearly up to $\sim\,200$ \kms\ in
the inner $8''$ ($3.0$ kpc), followed by a drop to $\sim\,170$ \kms\
between $8''$ and $15''$ on both sides of the nucleus; further out it
rises to $\sim\,240$ \kms\ at $20''$, and then remains constant. At
the center $\sigma_\star \sim\,210$ \kms, then off the nucleus it
decreases to $\sim\,50$ \kms.\\
{\it Gas.} $V_g$ has a steeper gradient than $V_\star$, peaking at 300
\kms\ at about $8''$; then it decreases, becoming $V_g
\sim V_\star$ at $20''$. The \ha\ shows steeper central RC gradient
and lower velocity dispersion than \nii\ and \sii (see
Fig.~\ref{fig:gas_b}).  This feature is probably due to the lower
$V_\star$: the \ha\ absorption does not have the same central
wavelength as the emission, and hence it shifts the resulting peak
toward higher rotation velocities.\\

\noindent
{\bf NGC~4698 --} The stellar and ionized-gas kinematics are measured
out to $75''$ ($4.4$ kpc) and $100''$ ($5.9$ kpc) on each side of the
nucleus, respectively.\\
{\it Stars.} In the innermost $10''$ ($0.6$ kpc) the stars have zero
rotation; at outer radii, $V_\star$ is less steep than $V_g$; only for
$|r| > 40''$ (2.3 kpc) are $V_\star \sim V_g$ and $\sigma_\star \sim
\sigma_g$. The profile of $\sigma_\star$ is radially asymmetric: in
the SE side it shows a maximum of $\sim\,150$ \kms\ at $9''$ ($0.5$
kpc), then it decreases outwards to $\sim\,50$ and 30
\kms\ at $30''$ ($1.8$ kpc) in the SE and NW sides, respectively. 
The measured $V_\star(0) \sim\,0$ \kms\ is explained by Bertola et
al. (1998b) as due to the presence of an orthogonal-rotating bulge.\\
{\it Gas.} $V_g$ increases to $\sim$130 \kms\ in the inner $18''$
($1.1$ kpc); then it increases more gradually reaching $\sim$200 \kms\
at $60''$ ($3.5$ kpc), and stays approximately constant farther out.
In the inner $\pm7''$ ($\pm0.4$ kpc, roughly coinciding with the
absorption lines region) $\sigma_g$ has a $75$ \kms\ plateau, while at
larger radii it drops to $\sigma_g \leq 50$ \kms.\\

\noindent
{\bf NGC~4845 --} The stellar and ionized-gas kinematics are measured
out to $70''$ ($4.4$ kpc) in the SW side, and out to $90''$ ($5.7$
kpc) in the NE side.\\
{\it Stars.} $V_\star$ has a shallower gradient than $V_g$: it reaches
60 \kms\ at $8''$ (0.5 kpc), and further out it increases slowly,
reaching the $V_g$ at $\sim\,60''$ (3.8 kpc). The velocity dispersion
is constant, $\sigma_\star \sim\,60$ \kms\ (in the SW side it drops to
30 \kms\ for $r > 40''$).\\
{\it Gas.} $V_g$ reaches $\sim$180 \kms\ at $14''$ (0.9 kpc), to
decrease and remain constant at 150 \kms\ farther out. For $|r| <
10''$ (0.6 kpc) $\sigma_g \simeq 80$ \kms, then it rapidly falls to
$<30$ and $\sim\,40$ \kms\ in the SW and NE sides, respectively;
farther out the behavior of $\sigma_g$ is more uncertain, due to a
considerable scatter of the measurements from different lines: along
the NW side $\sigma_g \sim\,40$ \kms, while along the SE side
$\sigma_g$ slowly decreases to $\sim\,30$ \kms .\\
The triaxiality of the bulge of NGC~4845 has been detected by Bertola,
Rubin \& Zeilinger (1989) and discussed by Gerhard, Vietri \& Kent
(1989).
\medskip

The RCs and velocity-dispersion profiles of both the ionized gas and
the stars in our Sa galaxies show a rich diversity of kinematical
properties.

$V_\star$ has shallower gradient than $V_g$ at the center, while
$V_\star = V_g$ at the last measured radius, in all our sample
galaxies. For NGC~2179, NGC~2775, NGC~3593, NGC~4698 and NGC~4845, the
gas RCs remain flat after a monotonic rise to a maximum [whose
observed values range between $\sim\,120$ \kms\ (as for NGC~3593) and
$\sim\,190$ \kms\ (as for NGC~2179, NGC~2775)], or rise monotonically
to the farthest observed radius (as for NGC~3281 and IC~724: in the latter
after an initial peak at $\sim\,290$ \kms). Stellar counterrotation
and orthogonal rotation has been found in NGC~3593 (Bertola et
al. 1996) and NGC~4698 (Bertola et al. 1998) respectively.

The observed $\sigma_\star$ exceeds 100 \kms\ for several kpc in the
innermost regions, peaking at values ranging between 130 \kms\ (as in
NGC~4698) and 210 \kms\ (as in IC~724); the only exception is NGC~4845
with $\sigma_\star \sim\,70$ \kms\ at all observed radii. There are
sample galaxies whose $\sigma_g$ is low at all radii, reaching a
central maximum of $\sim\,80$ \kms\ (NGC~4698, NGC~4845) or remaining
flattish at $\sim\,50$ \kms\ (NGC~3593); and others where
$\sigma_g$ increases to $> 100$ \kms\ either at the very center (as in
NGC~2179, NGC~2775) or over an extended radial range around the center
(as in NGC~3281, IC~724).

\section{Mass models}

Previous authors (Fillmore, Boroson \& Dressler 1986; Kent 1988;
Kormendy \& Westpfahl 1989) noticed that in the bulge of early-type
spirals $V_g$ falls below the predicted circular velocity. Such
`slowly rising' gas RCs are explained by Bertola et al. (1995) with
the argument that random (non-circular) motions are crucial for the
dynamical support of the ionized gas: in some galaxies of their S0
sample they measured $\sigma_g \sim \sigma_\star \ga150$ \kms\ over an
extended range of radii. We do observe the same phenomenon in some of
our early-type spirals (see Fig.~5): this fact prevents us to adopt,
for early-type disk galaxies, the inner portion of $V_g(r)$ as the
circular velocity on which to perform the mass decomposition.
When the high values for the velocity dispersion of the gas are
measured only in the very central parts (as in NGC~2179) we can not
exclude that this is an effect of rotational broadening due to the
seeing smearing of the steep velocity gradient.

At larger radii where $\sigma_g \la 50$ \kms, the ionized gas can be
considered a tracer of the actual circular velocity. But the limited
extension, ($0.5-1)\, R_{25}$ (see Tab.~1), of our $V_g(r)$ makes the
derivation of the halo parameters of early-type spirals more uncertain
than for later types. In fact, on one hand we lack data at very large
radii where only disk and halo affect the circular velocity, on the
other hand at small radii (where we can not consider the gas in
circular motion) not two (like for Sc-Sd galaxies), but three mass
components will have locally similar behaviors (solid-body like). So,
in absence of extended and complete RCs, an Sa mass solution would be
degenerate.

We therefore have to model the stellar kinematics to determine the
galaxy's total gravitational potential. Then we need to check the
derived mass decomposition by comparing the circular velocity,
inferred from the model, with $V_g$ at large galactocentric
radii. This is necessary to minimize uncertainties on the mass
structure obtained from the stellar kinematics. In fact, the uncertain
orbital structure of the spheroidal component, consistent with the
observed kinematics leads to a degeneracy between velocity anisotropy
and mass distribution, which can be solved only through the knowledge
of the line-of-sight velocity distribution profiles (Gerhard 1993).

Two galaxies of the observed sample, namely NGC~2179 and NGC~2775, are
particularly suited to be studied with the three-component mass
models, based on stellar photometry and kinematics, at our disposal.
They were chosen for their nearly axisymmetric stellar pattern and to
not contain kinematically decoupled (as found NGC~3593 and NGC~4698) 
or triaxial (as found in NGC~4845) stellar components.

\subsection{The modeling technique}

We apply the Jeans modeling technique introduced by Binney, Davies \&
Illingworth (1990), developed by van der Marel, Binney \& Davies
(1990) and van der Marel (1991), and extended to two-component
galaxies by Cinzano \& van der Marel (1994) and to galaxies with a DM
halo by Cinzano (1995).  (For a detailed description of the model and
its assumptions, see the above references.)

The galaxy is assumed to be axisymmetric. Its mass structure results
from the contributions of: {\it (i)} a spheroidal component; {\it
(ii)} an infinitesimally thin exponential disk; and {\it (iii)} a
spherical pseudo-isothermal dark halo with density distribution
$\rho(r)=\rho_0/\left[1+(r/r_h)^2\right]$. The mass contribution of
the ionized gas is assumed to be negligible at all radii.  The
spheroidal and disk components are supposed to have constant $M/L$
ratios.  The total potential is the sum of the (numerically derived)
potential of the spheroid plus the (analytical) potentials of the disk
and the halo.

The stellar distribution function $f$ is assumed to depend only on two
integral of motion [i.e., $f=f(E,L_z)$]. In these hypotheses the Jeans
equations for hydrostatic equilibrium form a closed set that, once
solved in the total potential, yields the dynamical quantities to be
compared with the observed kinematics, once projected onto the sky
plane.

To obtain the potentials of the bulge and the disk, we proceed through
several steps. {\it (a)} First, the bulge surface brightness is
derived from the total one by subtracting the disk.  Then, it is
deprojected by means of Lucy's algorithm to yield the 3-D luminosity
density which, via the $M/L$ ratio, gives the 3-D mass density of the
bulge. Finally, solving Poisson's equation through multipole
expansion, we derive the bulge potential (Binney et al. 1990). {\it
(b)} The exponential disk parameters (scale length $r_d$, central
surface brightness $\mu_0$, and inclination $i$) are chosen according
to the best-fit photometric decomposition.  If $\overline{r_d}$,
$\overline{\mu_0}$ and $\overline{i}$ are the disk parameters
resulting from the photometric decomposition, the best-fit model to
the observed stellar kinematics was obtained considering exponential
disks with $|r_d-2''| \leq \overline{r_d}$, $|\mu_0-0.3|$ \maq\ $\leq
\overline{\mu_0}$, and $|i-5\degr| \leq \overline{i}$.
These parameters determine the surface brightness of the disk.
Through the disk $M/L$ ratio we obtain the surface mass density of the
disk, and then its potential (Binney \& Tremaine 1987).

We first solved the Jeans equations only for both the bulge and disk
components in their total potential, to give in every point of the
galaxy the velocity dispersions onto the meridional plane
$\sigma^2_R=\sigma_z^2$ and the mean azimuthal squared velocities
$\overline{v_{\phi}^2}$.

To disentangle the respective contributions of the azimuthal velocity
dispersion $\sigma^2_{\phi}$ and the mean stellar motion
$\overline{v}^2_{\phi}$ to $\overline{v_{\phi}^2}$, for the bulge we
made the same hypotheses of Binney et al. (1990) while for the disk we
followed Cinzano \& van der Marel (1994) respectively.  Part of the
second azimuthal velocity moment $\overline{v^2_{\phi}}$ in the bulge
is assigned to the streaming velocity $\overline{v_\phi}$ as in Satoh
(1980).  The azimuthal velocity dispersion $\sigma_{\phi}^2$ in the
disk is assumed to be related to $\sigma_{R}^2$ (which is assumed in
turn to have an exponential fall-off with central value
$\sigma_{R,0}^2$ and scale-length $r_\sigma$) according to the
epicyclic theory (cfr. Binney \& Tremaine 1987).

In the framework of Cinzano \& van der Marel (1994), we have to take into
account the effects of seeing, of finite slit-width and pixel-size in
data acquisition, and of Fourier filtering in data reduction (notably 
the wavenumber range, as discussed in Sect.~2.1.), in order to compare the 
sky-projected model predictions with the observed stellar kinematics.

We interpret the discrepancy between the model's circular velocity and
the observed gas rotation in the outer regions as due to the presence
of a DM halo. In this case, the Jeans equations have to be solved
again, taking the halo into account, too. By introducing the DM halo,
the number of free parameters of the model increases to ten.  They
are: $k$, the local rotation anisotropy parameter of the bulge; the
$M/L$ ratios of the bulge and the disk; the disk central surface
brightness, scale length and inclination; the central value and scale
length of the disk's second radial velocity moment; and the halo's
central mass density and core radius.  To reduce the number of free
parameters, in the following we consider only three-component models
having same best-fit parameters as no-halo models except for the bulge
and disk $M/L$ ratios.  We choose the fit parameters in order to
simultaneously reproduce the stellar kinematics at all radii as well
as $V_g$ at large radii.

The modeling technique described above derives the 3-D distribution of
the luminous mass from the 3-D luminosity distribution inferred from
the observed surface photometry.  For this reason, in the central
regions we take into account the seeing effects on the measured
photometrical quantities (surface brightness, ellipticity and $\cos
4\theta$ deviation profiles).

We derive for NGC~2179 and NGC~2775 the seeing cutoffs $r_{\mu}$ and
$r_{\epsilon}$, defined by Peletier et al. (1990) as the radii beyond
which the seeing-induced error on the profile is lower than,
respectively, 0.05 \maq\ in surface brightness and 0.02 in
ellipticity.  They have expressed $r_{\mu}$ and $r_{\epsilon}$ for a
de Vaucouleurs profile as a function of the seeing FWHM, the effective
radius $r_e$ and the ellipticity $\epsilon$. For the bulge, we
obtained $r_e$ and $\epsilon$, and the corresponding seeing cutoffs
$r_{\mu}$ and $r_{\epsilon}$, following an iterative procedure.  We
started by performing a standard bulge-disk decomposition with a
parametric fit (e.g. Kent 1985): we decomposed the observed
surface-brightness profile on both the major and the minor axis as the
sum of a de Vaucouleurs bulge of surface-brightness profile

\begin{equation}
\mu_b = \mu_e + 8.3268 \left[\left(\frac{r}{r_e}\right)^{1/4}-1 \right]\,,
\end{equation}
plus an exponential disk of surface-brightness profile
\begin{equation}
\mu_d = \mu_0 + 1.0857 \left(\frac{r}{r_d}\right)\,.
\end{equation}

We assumed the minor-axis profiles of each component to be the same as
the major-axis profiles, with values scaled by a factor $1-\epsilon =
b/a$.  A least-squares fit of the model to the photometric data
provided $r_e$, $\mu_e$ and $\epsilon$ of the bulge, $\mu_0$, $r_d$ of
the disk, and the galaxy inclination $i$. The values of $r_e$ and
$\epsilon$ were used as a starting input to derive $r_{\mu}$ and
$r_{\epsilon}$. Following van der Marel (1991), the ellipticity
$\epsilon$ and the $\cos 4\theta$ Fourier coefficients were kept
constant within $r_{\epsilon}$ to their value at $r_{\epsilon}$, and
the surface-brightness profile was truncated at its value at
$r_{\mu}$.  A new parametric bulge-disk decomposition was then
performed on the truncated photometric data. The resulting new values
of the effective radius and ellipticity of the bulge were in turn used
to obtain a further estimate of $r_{\mu}$ and $r_{\epsilon}$. The
surface photometry was again modified according to these new values,
and then another parametric fitting was done.  The process was
repeated up to convergence.

A least-squares fit to $\mu(r)$ in the bulge-dominated region beyond
$r_{\mu}$ was performed using the 2-D brightness distribution
resulting from the 3-D luminosity density given for a spherical body
by:

\smallskip

\noindent
{\it (i)} a modified Hubble law (Rood et al. 1972):
\begin{equation}
j_{\it hu}(r) = j_0 
\left[1+\left(\frac{r}{a_{\it hu}}\right)^2\right]^{-\frac{3}{2}}
\label{eq:hubble}
\end{equation}
where $j_0$ and $a_{\it hu}$ are respectively the central luminosity 
density and the core radius;

\smallskip

\noindent
{\it (ii)} a Jaffe law (Jaffe 1983):
\begin{equation}
j_{\it ja}(r) = \frac{L_{\it tot}}{4 \pi a_{\it ja}^3}
\left(\frac{a_{\it ja}}{r}\right)^2 \frac{1}{(1+r/a_{\it ja})^2}
\label{eq:jaffe}
\end{equation}
where $L_{\it tot}$ and $a_{\it ja}$ are the total luminosity and the 
half-light radius;

\smallskip

\noindent
{\it (iii)} a Hernquist law (Hernquist 1990):
\begin{equation}
j_{\it he}(r) = \frac{L_{\it tot}}{2 \pi} \frac{a_{\it he}}{r} 
\frac{1}{(r + a_{\it he})^3}
\label{eq:hernquist}
\end{equation}
where $L_{\it tot}$ and $a_{\it he}$ are the total luminosity and a 
scale radius.

\smallskip

The best fit was achieved for NGC~2179 with a Hernquist profile and
for NGC~2775 with a modified Hubble profile.  We used them to
extrapolate the $\mu(r)$ profiles of the two galaxies to $r <
r_{\mu}$.  After subtracting the disk contribution from the total
surface brightness, the 3-D luminosity density of the bulge was
obtained starting Lucy's iterations from a flattened Hernquist model
for NGC~2179 and from a flattened modified Hubble model for
NGC~2775. The radial profiles of these flattened models are derived
respectively from Eqs.~\ref{eq:hernquist} and \ref{eq:hubble} by
replacing $r$ with $(q^2R^2+z^2)^{1/2}$ where $q$ is the flattening
and $R,z$ cylindrical coordinates.

\subsection{The modeling results}

In this section we present the mass models of NGC~2179 and NGC~2775.

\subsubsection{NGC~2179}

\begin{figure}[ht] 
\vspace*{11cm}
\includegraphics{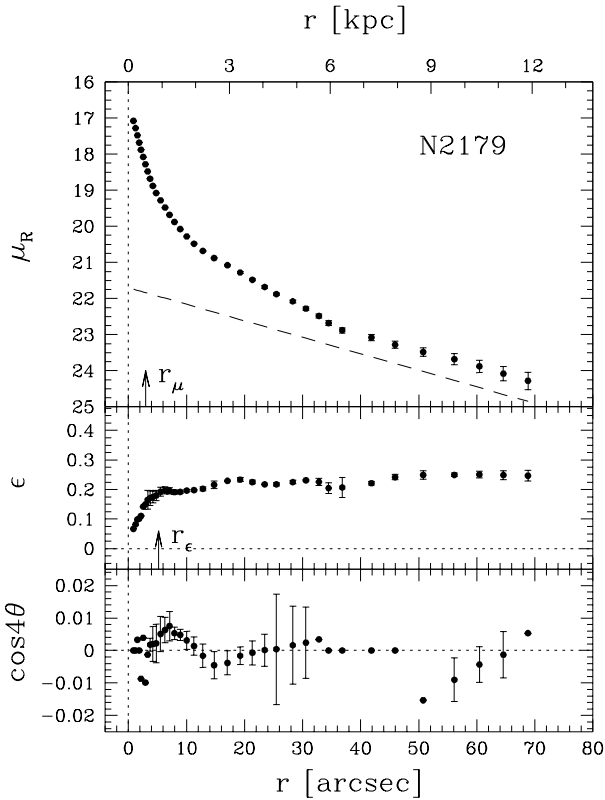}
\vspace*{.25cm}
\caption[]{The NGC~2179 $R-$band surface brightness, ellipticity and
  $\cos 4\theta$ coefficient profiles as a function of radius along the
  major axis. The arrows in the surface brightness and in the
  ellipticity panels indicate the position of the seeing cutoff radii
  $r_\mu$ and $r_\epsilon$.  The {\it dashed curve\/} in the upper
  panel is the surface brightness exponential profile of the disk
  component of the best fit model to the stellar kinematics}
\label{fig:n2179fot}
\end{figure}

In Fig.~\ref{fig:n2179fot} 
we show the $R$-band surface brightness ($\mu_R$), ellipticity
($\epsilon$) and $\cos 4\theta$ Fourier coefficient of the isophote
deviations from elliptical, as a function of radius along the major
axis.

The seeing cutoffs are $r_{\mu}=3\farcs0$ and $r_{\epsilon}=5\farcs2$.
The corresponding best-fit parameters obtained from the photometric
decomposition are: $\mu_e = 21.0$ $R-$\maq, $r_e = 12\farcs4$,
$\epsilon_b= 0.58$ for the de Vaucouleurs bulge; $\mu_0 = 21.8$
$R-$\maq, $r_d = 23\farcs5$, $\epsilon_d=0.29$ for the exponential
disk. (Taking into account the photometric bulge-disk decomposition,
the exponential disk yielding the best-fit model to the observed
stellar kinematics has $\mu_0 = 21.7$ $R-$\maq , $r_d = 23\farcs7 =
4.1$ kpc, and $i=45\degr$; see dashed curve in
Fig.~\ref{fig:n2179fot}). We then subtract the disk contribution from
the total surface brightness. The residual surface brightness is the
contribution of the spheroidal component.

\begin{figure}[ht] 
\vspace*{6cm}
\includegraphics{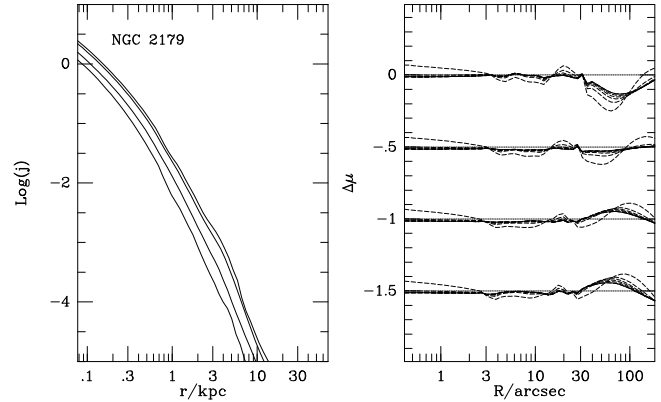}
\vspace*{.25cm}
\caption[]{The deprojection of the surface brightness of the
  spheroidal component of NGC~2179.  The {\it right panel\/} shows the
  difference $\mu_{\rm model} - \mu_{\rm obs}$ after each Lucy
  iteration ({\it dashed lines\/}) starting from an initial Hernquist
  fit to the actual NGC~2179 bulge brightness.  The residuals are shown
  for four axes (major through minor axis: top through bottom).  For
  each set of curves: the {\it solid line\/} corresponds to the
  projected adopted model for 3-D luminosity density; and the {\it
  dotted line\/} corresponds to a perfect deprojection.  At each
  iteration if the model is brighter than the galaxy, the
  corresponding {\it dashed\/} or {\it continuous\/} curve is below the
  {\it dotted line\/}. In the {\it left panel} the final 3-D
  luminosity density profile of the spheroidal components of NGC~2179 is
  given in units of $10^{10} {\rm L_{\odot}\;kpc^{-3}}$ for the same
  four axes (minor through major axis: innermost through outermost
  curve)}
\label{fig:n2179dep}
\end{figure}

The difference between the surface brightness of the sphe\-roid and
that obtained projecting the 3-D luminosity distribution of each of
the four Lucy iterations (including the initial flatted Hernquist
model) is shown in Fig.~\ref{fig:n2179dep} (right panel) along
NGC~2179's major, minor and two intermediate axes. (The
r.m.s. residual of the last Lucy iteration corresponds to 0.06509 \maq
). The 3-D luminosity density of the final bulge model along the same
four axes is also presented in Fig.~\ref{fig:n2179dep} (left panel).

We fold $v_g$ and $v_\star$ around their respective centers of
symmetry.  In order to determine the latter, we fit a suitable odd
function to both RCs independently: this yields the position of the
kinematical center of the curve, $r_0$, and the heliocentric velocity
of the galaxy, $V_\odot$.  We find $V_\odot = 2885\pm10$ \kms\ for
both gas and stars, and $r_{0, g}=+0\farcs5\pm0\farcs3$ and
$r_{0,\star}=+0\farcs6\pm0\farcs3$. We then fold $\sigma_g$ and
$\sigma_\star$ around the kinematical center of the respective
component.

\begin{figure}[ht] 
\vspace*{6cm}
\includegraphics{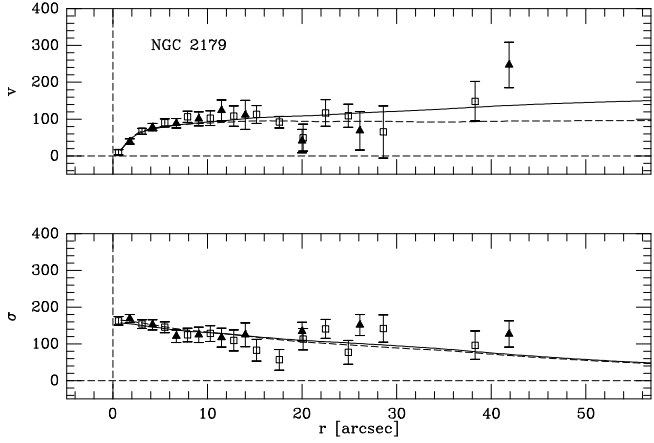}
\vspace*{.25cm}
\caption[]{Predicted and observed stellar kinematics of NGC~2179.  The
  {\it filled triangles\/} and the  {\it open squares\/} represent data
  derived for the approaching SE side and for the receding NW side
  respectively.  The {\it continuous\/} and the {\it dashed lines\/} are
  the model stellar velocities ({\it upper panel\/}) and velocity
  dispersions ({\it lower panel\/}) obtained with and without the dark
  halo component respectively}
\label{fig:n2179mod_s}
\end{figure}

The best-fit model to the observed major-axis stellar kinematics is
shown in Fig.~\ref{fig:n2179mod_s}.  Its parameters are as follows.
The bulge is an oblate isotropic rotator ($k=1$) with $(M/L_R)_b$ =
6.1 \mlR . The exponential disk has $\sigma_\star(r)=
168\;e^{-r/r_\sigma}$ \kms\ with scale-length $r_{\sigma} = 23\farcs2
= 4.0$ kpc), and $(M/L_R)_d$ = 6.1 \mlR .  The derived bulge and disk
masses are $M_{b}= 7.0 \cdot 10^{10}$ \msun\ and $M_{d}= 2.5 \cdot
10^{10}$ \msun , adding up to a total (bulge + disk) luminous mass of
$M_{\rm LM} = 9.5 \cdot 10^{10}$ \msun .  The DM halo has $\rho_0 =
6.9 \cdot 10^{-2}$ \msunpc\ and $r_h = 24'' = 4.2$ kpc, which
correspond to an asymptotic rotation velocity $V_\infty = 257$ \kms;
its mass at the outermost observed radius is $M_{\rm DM}=6.5 \cdot
10^{10}$ \msun.  The ratio between the mass-to-light ratios of the
stellar components in the models with and without the DM halo is 0.9.

\begin{figure}[ht] 
\vspace*{6cm}
\includegraphics{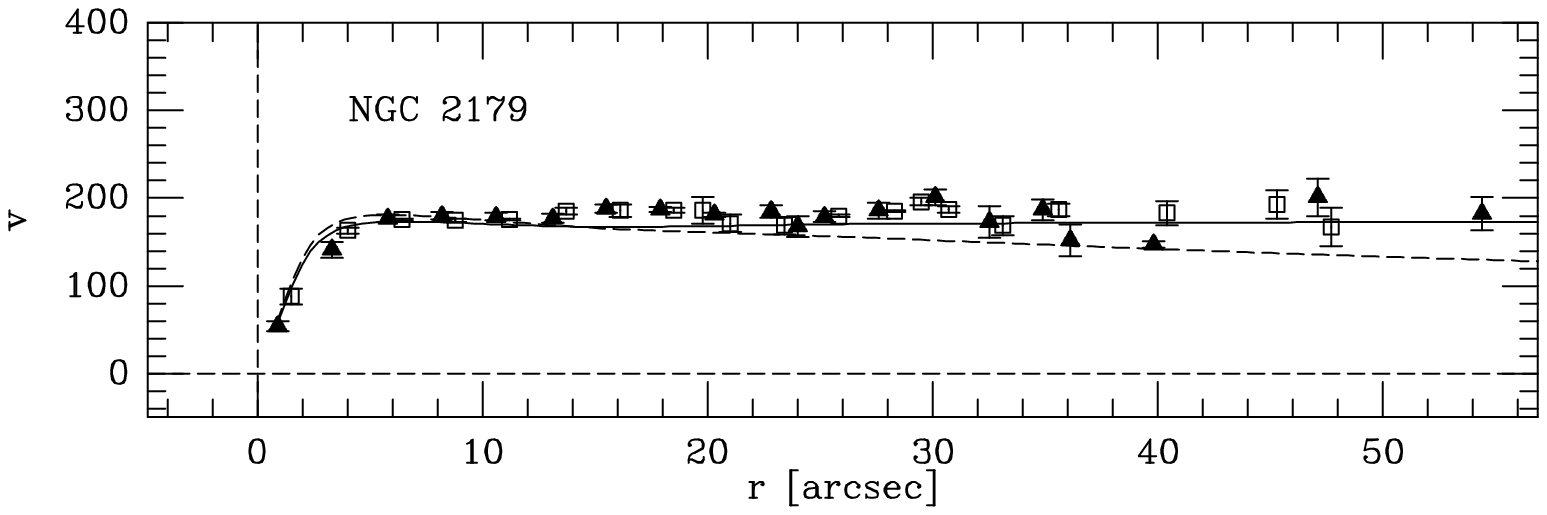}
\includegraphics{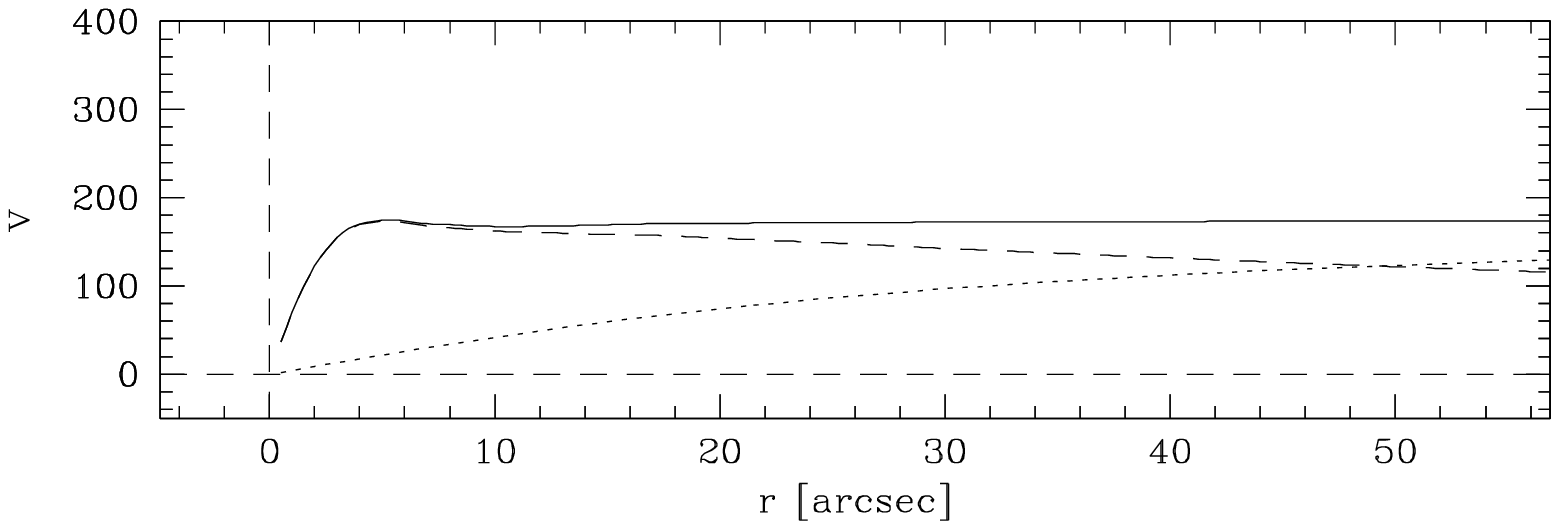}
\vspace*{.25cm}
\caption[]{In the {\it upper panel\/} the predicted circular velocity 
  and the observed gas rotation velocity of NGC~2179 are shown. The
  {\it filled triangles\/} and the {\it open squares\/} represent data
  derived for the approaching side and for the receding side
  respectively.  The {\it continuous\/} and the {\it dashed lines\/}
  are the total circular velocities derived from the stellar
  kinematics with and without the dark halo component respectively.
  In the {\it lower panel\/} the maximum contribution of the luminous
  matter ({\it dashed line\/}) to the total circular velocity ({\it
  continuous line\/}) is plotted if a dark matter halo 
  ({\it dotted line\/}) is considered}
\label{fig:n2179mod_g}
\end{figure}

The comparison between the observed rotation of the ionized gas and
the true circular velocity, inferred from stellar kinematics, is given
in the upper panel Fig.~\ref{fig:n2179mod_g}.  It shows that a DM halo
is unambiguously required to explain the rotation at large radii
($r\geq25''$).  This result hinges on accuracy of the gas kinematics
data beyond $40''$ mostly derived from \ha\ line measurements.  In
NGC~2179, the gas rotation does provide the circular velocity at all
radii. The contribution of the DM halo to NGC~2179 circular velocity
as function of radius is plotted in lower panel of
Fig.~\ref{fig:n2179mod_g}.

\subsubsection{NGC~2775}

\begin{figure}[ht] 
\vspace*{9.5cm}
\includegraphics{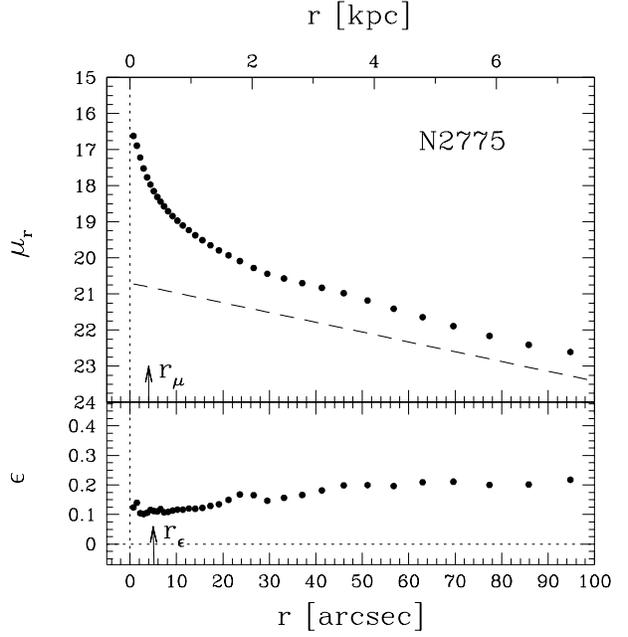}
\vspace*{.25cm}
\caption[]{The NGC~2775 $r-$band surface brightness and the ellipticity
  profiles as a function of radius along the major axis (Kent 1988).
  The {\it dashed curve\/} in the {\it top panel\/} is the surface
  brightness exponential profile of the disk component of the best fit
  model to the stellar kinematics}
\label{fig:n2775fot}
\end{figure}

In Fig.~\ref{fig:n2775fot} we show $\mu_r(r)$ and $\epsilon(r)$ along
the major axis (Kent 1988).  The seeing FWHM for the Kent (1988) data
is $2\farcs3$.  We derive the seeing cutoffs $r_{\mu}=4\farcs0$ and
$r_{\epsilon}=5\farcs2$.  As no $\cos 4\theta$ profile is available,
we assume it is zero throughout.

The photometric model is improved by taking into account an outer dust
lane surrounding the spiral pattern of the galaxy. (This appears as a
thin dust ring at about $80''$ from the center, and is visible on
panels 78 and 87 of the CAG, see Sandage \& Bedke 1994). As a fitting
function for the disk component, we use an exponential profile
weighted by an absorption ring.  We assume the section of the dust
ring to have a Gaussian radial profile, defined by central intensity
maximum absorption $A_r$, a center $r_r$ and a scale-length
$\sigma_r$. This leads to a disk light contribution 

\begin{equation}
I_{\it d,ring}(r)=I_d(r)\;\left(1-A_r\;
e^{-\left(\frac{r-r_r}{\sigma_r}\right)^2}\right)
\end{equation} 

The best-fit parameters resulting from the improved photometric
decomposition are: $\mu_e = 22.0$ $r$-\maq, $r_e = 53\farcs0$,
$\epsilon_b=0.10$ for the de Vaucouleurs bulge; $\mu_0 = 20.7$
$r$-\maq, $r_d = 40\farcs5$, $\epsilon_d=0.28$ for the exponential
disk; and $A_r=0.25$, $r_r=85''$ and $\sigma_r= 20''$ for the dust
ring.  The exponential disk yielding to the best-fit model to the
observed stellar kinematics has $\mu_0 = 20.7$ $r$-\maq, $r_d =
40\farcs0 = 3.0$ kpc, and $i=44\degr$ (see Fig.~\ref{fig:n2775fot}).

\begin{figure}[ht] 
\vspace*{6cm}
\includegraphics{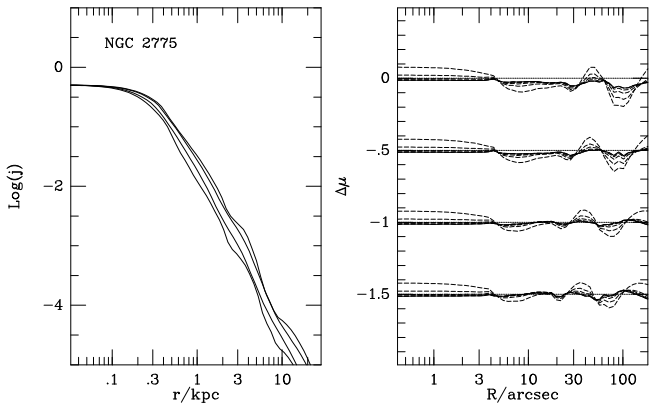}
\vspace*{.25cm}
\caption[]{Same as Fig.~7, but for NGC~2775 and starting from 
  a flattened modified Hubble model for the spheroidal component}
\label{fig:n2775dep}
\end{figure}

The surface brightness of the exponential disk in Fig.~\ref{fig:n2775fot} 
is subtracted from the total surface brightness.  The fit to the
spheroidal component's deprojected surface brightness is obtained
after four Lucy iterations from an initial flatted modified Hubble
model (The r.m.s. residual is 0.03209 \maq , see Fig.~\ref{fig:n2775dep}). 
The 3-D luminosity density of the final bulge model along the major,
minor and two intermediate axes is also shown in Fig.~\ref{fig:n2775dep}.

We find $v_g$ and $v_\star$ to have same center of symmetry at
$r_{0,g}=r_{0,\star}=+4\farcs0\pm0\farcs3$ and $V_\odot = 1350\pm10$
\kms.  The velocity dispersion profiles are folded around the
kinematical center.

\begin{figure}[ht] 
\vspace*{6.5cm}
\includegraphics{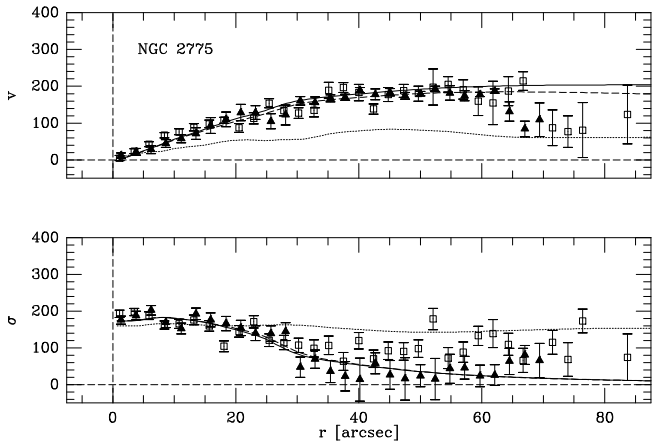}
\vspace*{.25cm}
\caption[]{Same as Fig.~8, but for NGC~2775.  The {\it filled
  triangles\/} and the {\it open squares\/} represent data derived for
  the receding SE side and for the approaching NW side respectively.
  The {\it dotted line\/} is the model kinematics for the
  spheroidal component}
\label{fig:n2775mod_s}
\end{figure}

The best-fit model to the observed major-axis stellar kinematics is
shown in Fig.~\ref{fig:n2775mod_s}.  Its parameters are as follows.
The bulge is an oblate isotropic rotator ($k=1$) with $(M/L_r)_b$ =
5.2 \mlr .  The exponential disk has $\sigma_\star(r)=
130\;e^{-r/r_\sigma}$ \kms\ with scale-length $r_{\sigma} = 17\farcs5
= 1.3$ kpc), and $(M/L_r)_d$ = 7.0 \mlr . The derived bulge and disk
masses are $M_{b}= 8.5 \cdot 10^{10}$ \msun and $M_{d}= 6.1 \cdot
10^{10}$ \msun , so the total (bulge + disk) luminous mass is $M_{\rm
LM} = 14.6 \cdot 10^{10}$ \msun .  We kept the ratio between the bulge
and disk $M/L$ fixed at the value 1.36 (see Kent 1988).  The DM halo
has central density $\rho_0 = 5.8 \cdot 10^{-2}$ \msunpc\ and core
radius $r_h = 60'' = 4.6$ kpc, which correspond to an asymptotic
rotation velocity $V_\infty = 258$ \kms; its mass at the outermost
observed radius is $M_{\rm DM}=3.1 \cdot 10^{10}$ \msun.  (The latter
should be considered an upper limit because the inner kinematics can
be explained with no DM halo).  Contrary to NGC~2179 in this case we
kept the same mass-to-light ratios of the stellar components in the
models with and without the DM halo.

For $|r|>65''$ the presence on the equatorial plane of the dust ring,
which reduces the light contribution of the (faster-rotating) disk
stars of a further $\sim25\%$, causes the observed stellar kinematics
to be more affected by the (slower-rotating) bulge stars.  In this
picture as shown in Fig.~\ref{fig:n2775mod_s}, our bulge model agrees
with the observed drop in velocity and the rise in velocity
dispersion. NGC~2775 is a dust-rich system, as can be inferred from
its dust-to-HI mass ratio (Roberts et al. 1991), which is $\sim12$
times larger than the mean S0/Sa values (Bregman, Hogg \& Roberts
1992).

Although the total luminous mass found by Kent (1988) $M_{\rm LM} =
14.3 \cdot 10^{10}$ \msun\ (with $H_0=75$ \kms\ Mpc$^{-1}$) is in good
agreement with ours, his mass decomposition differs from ours.  He
assumed only the bulge to have an analytical $\mu(r)$ (a de
Vaucouleurs law with $\mu_e = 21.0$ \maq, $r_e = 22\farcs9$,
$\epsilon_b=0.12$), while the disk $\mu(r)$ was taken to be the
bulge-subtracted major-axis profile. Kent's approach is opposite to
ours.  We assume the disk to have an analytical $\mu(r)$, and the
bulge $\mu(r)$ to be the residual surface brightness after subtracting
the disk (assuming no a priori analytical expression or fixed axis
ratio). Scaled to our assumed distance, the luminosity of Kent's bulge
is $\sim\,31\%$ of ours, while our disk luminosity is $\sim\,43\%$ of
Kent's.

\begin{figure}[ht] 
\vspace*{6cm}
\includegraphics{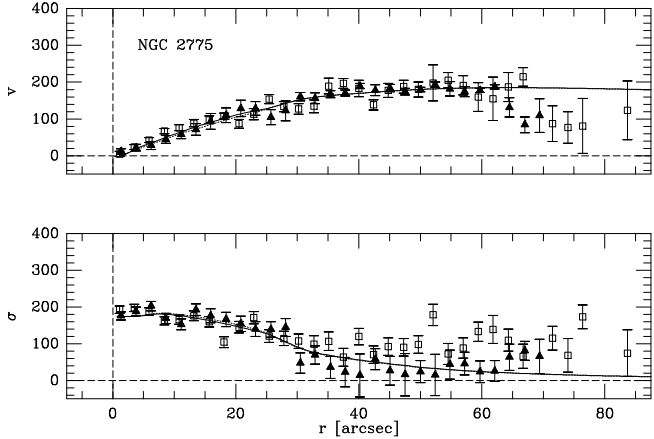}
\vspace*{.25cm}
\caption[]{Predicted stellar kinematics for NGC~2775 with fixed 
  $\cos 4\theta=+0.02$ ({\it solid line}) and fixed $\cos 4\theta=-0.02$ 
({\it dotted line}) at all radii} 
\label{fig:n2775mod_a4}
\end{figure}

Van der Marel et al. (1991) studied the effects of a $\cos 4\theta$
deviation on the kinematics of NGC~4261. They found that changing the
$\cos 4\theta$ Fourier component from zero to $\pm 0.02$ produces
variations of $\sim 2\%$ in velocity dispersion and $\leq 10\%$ in
rotation velocity.  Fig.~\ref{fig:n2775mod_a4} shows the stellar
kinematics of NGC~2775 in the case of slightly disky ($\cos 4\theta
=+0.02$) and slightly boxy ($\cos 4\theta = -0.02$) isophotes (solid
and dotted line, respectively).  The disky model rotates faster and
has a lower $\sigma_\star$ than the boxy model in the inner $25''$
(1.9 kpc).  The differences in $V_\star$ and in $\sigma_\star$ between
the two models are $< 10$ \kms : this means that, in the observed
range of values, a difference of $0.04$ in $\cos 4\theta$ coefficients
corresponds to a difference of $<12\%$ in velocities and $<7\%$ in
velocity dispersions.  However, these uncertainties are immaterial to
our results on the mass structure of NGC~2775 .

\begin{figure}[ht] 
\vspace*{6cm}
\includegraphics{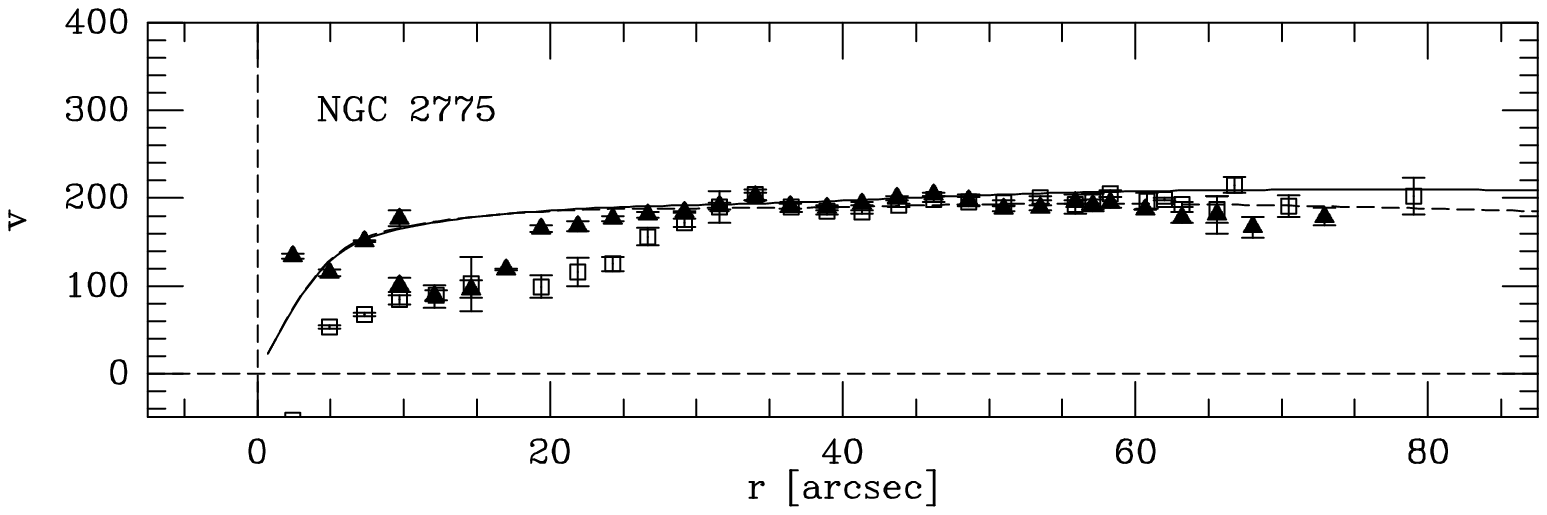}
\includegraphics{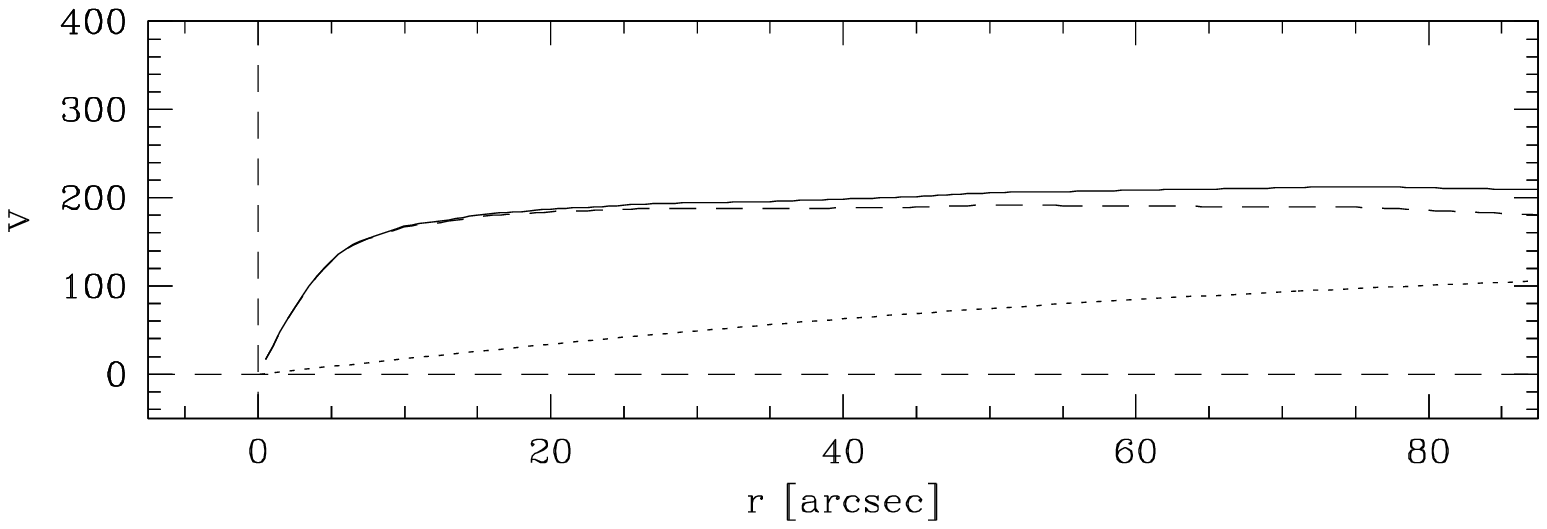}
\vspace*{.25cm}
\caption[]{Same as Fig.~9, but for NGC~2775.  The {\it filled
  triangles\/} and the {\it open squares\/} represent data derived for
  the receding SE side and for the approaching NW side respectively}
\label{fig:n2775mod_g}
\end{figure}

The comparison between $V_g$ and the true circular velocity inferred
from the stellar kinematics is shown in the upper panel of
Fig.~\ref{fig:n2775mod_g}.  It shows that a DM halo is not strictly
required to explain the rotation at large radii ($r \geq 35''$). The
contribution of the DM halo to NGC~2775 circular velocity as a
function of radius is plotted in the lower panel of
Fig.~\ref{fig:n2775mod_g}.  Inside $30''$ on the receding arm,
$V_\star \simeq V_g$ and $\sigma_\star > \sigma_g \simeq 50$
\kms. This rules out the case that the gas kinematics is dominated by
random motions, and leads us to speculate that we are looking at gas
rotating on a non-equatorial plane. We suggest this is the signature
of a past external acquisition (possibly from the companion galaxy
NGC~2777) of gas still not completely settled onto the disk plane.

\section{Discussion and conclusions}

We have presented the ionized-gas and stellar kinematics, measured
along the major axis, for a sample of six early-type spiral
galaxies. (Due to the high values of $\sigma_g$ in their inner
regions, the gas RCs can not be used as circular-velocity curves.)

For NGC~2179 and NGC~2775, we have modeled both the stellar and the
gaseous kinematics to derive the mass contribution of the luminous and
dark matter to the total potential, improving on the efforts by Kent
(1988) from gas kinematics alone.
 
We have found that the innermost kinematics ($r < 2 R_D$) is very well
and uniquely reproduced by taking into account the two luminous
components. In the (very luminous) early-type spirals considered here,
there is a large inner region in which (essentially) light traces the
mass and the DM is a minor mass component. This is agreement with the
`weak' maximum disk paradigm proposed by Persic \& Salucci (1990), but
in disagreement with the claim by Courteau \& Rix (1998) according to
which in the most luminous spirals DM is a protagonist at essentially
any radii.

More in detail we have found that in NGC~2179 the combined stellar and
gaseous rotation data (measured out to about $R_{\rm opt}$) require
the presence of a massive dark halo. In NGC~2775, more luminous and
massive than NGC~2179, we can rule out a significant halo contribution
out to $0.6\,R_{\rm opt}$.  This result complies with the general
trend of mass distribution known for later spirals (Persic et
al. 1996).

\begin{figure}[ht] 
\vspace*{8.5cm}
\includegraphics{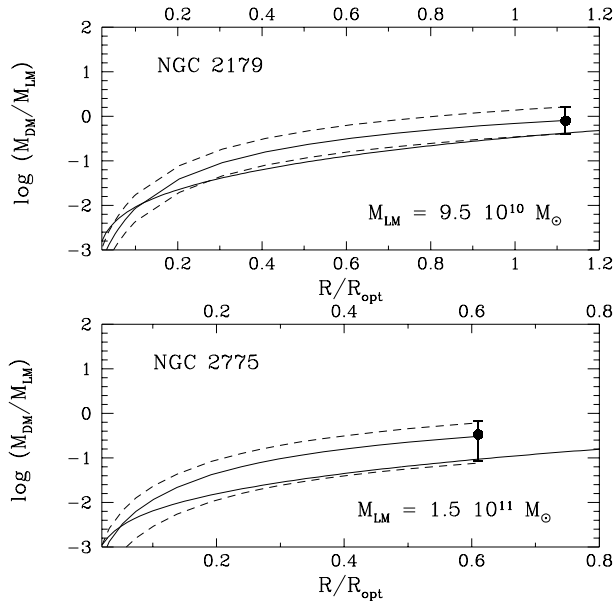}
\vspace*{.25cm}
\caption[]{The dark-to-luminous mass ratio as a function of radius
(normalized to $R_{\rm opt}$) for late spirals of same stellar mass as
NGC~2179 and NGC~2775 ({\it solid lines}), compared with our derived
values for NGC~2179 and NGC~2775 ({\it filled circles})}
\label{fig:dm}
\end{figure}

Salucci \& Persic (1997), considering a large number of galaxies of
mixed morphologies (ellipticals, late spirals, dwar\-fs, and LSBs), have
suggested that the halo structural parameters and the connection
between the dark and the luminous matter show a strong continuity when
passing from one Hubble Type to another. Ellipticals, considered as
luminous spheroids, and spirals, considered as luminous disks, are
evidently very different systems, markedly discontinuous in terms of
the distribution and global properties of the luminous matter.
However, in the structural parameter space, ellipticals and spirals
are contiguous, the main difference being that the former are more
concentrated in both the dark and luminous components, probably due to
the baryons' dissipational infall being deeper in ellipticals than in
spirals (e.g., Bertola et al. 1993). If so, it is hardly surprising
that Sa galaxies, being in some sense intermediate systems consisting
of a luminous spheroid embedded in a luminous disk, fit in the
regularity pattern of the dark-to-visible mass connection shared by
ellipticals and spirals. In Fig.~\ref{fig:dm} 
we plot our derived dark-to-visible mass ratios
at the farthest measured radii for NGC~2179 and NGC~2775 (filled circles)
onto the distribution, derived by Salucci \& Persic (1997) for
galaxies with the same visible mass. The agreement is good. 
Even if the present result on early-type spirals is preliminary
and without pretending to draw general conclusions from one particular
case, it nevertheless seems to agree with the idea that, for galaxies
of all morphological types, the dark-to-luminous mass ratio at
any given radius depends only on the (luminous) mass of the galaxy.

\acknowledgements

We are indebted to R.P. van der Marel for providing his $f(E,L_z)$ modeling 
software which became the basis of our modeling package. We also thank 
R. Falomo for providing some photometric data reduction tools. We are most 
grateful to the Vatican Observatory Research Group for allocating the 
observing time. Particular thanks go to R. Boyle, S.J. for his help during 
the observing run at the VATT. The research of AP was partially supported
by an {\em Acciaierie Beltrame\/} grant. 
JCVB acknowledges a grant from Telescopio Nazionale Galileo and 
Osservatorio Astronomico di Padova.
This research has made use of the NASA/IPAC Extragalactic Database
(NED) which is operated by the Jet Propulsion Laboratory, California
Institute of Technology, under contract with the National Aeronautics
and Space Administration.

\end{document}